\documentclass[12pt]{article}
\usepackage[T2A]{fontenc}
\usepackage{graphicx}
\usepackage[cp1251]{inputenc}
\begin{document}
\newcommand{\iso}[2]{\mbox{$^{#1}{\rm #2}$}}
\newcommand{\Teff}{T_{\rm eff}}
\newcommand{\kms}{km\,s$^{-1}$}
\newcommand{\kH}{$S_{\!\rm H}$}    
\newcommand{\eps}[1]{\log\varepsilon_{\rm #1}}
\newcommand{\ba}[4]{\mbox{$#1 ^2{\rm #2}^{#3}_{\rm #4}$}}
\newcommand{\Vmac}{V_{\rm mac}}

\centerline{\bf Influence of Inelastic Collisions with Hydrogen Atoms}
\centerline{\bf on the Formation of Al~I and Si~I Lines in Stellar Spectra}

\bigskip
\centerline{Lyudmila Mashonkina$^*$\footnote{\tt E-mail: lima@inasan.ru}, Andrey K. Belyaev$^{**}$\footnote{\tt E-mail: andrey.k.belyaev@gmail.com}, Jianrong Shi$^{***}$}

\bigskip
\centerline{\it $^*$Institute of Astronomy, Russian Academy of Sciences,}
\centerline{\it Pyatnitskaya st. 48, 119017 Moscow, Russia}
 
\centerline{\it $^{**}$Department of Theoretical Physics and Astronomy, Herzen University,}
\centerline{\it Moika st. 48, St. Petersburg 191186, Russia }

\centerline{\it $^{***}$National Astronomical Observatories, Chinese Academy of Sciences,}
\centerline{\it  A20 Datun Road, Chaoyang District, Beijing, 100012 China }

\bigskip
{\bf Abstract -} The non-local thermodynamic equilibrium (non-LTE) line formation for Al~I and Si~I was calculated with model atmospheres corresponding to F-G-K type stars of different metal abundances. To take into account inelastic collisions with neutral hydrogen atoms, for the first time we have applied the cross sections calculated by Belyaev et al. using 
model approaches within the formalism of the Born-Oppenheimer quantum theory. We show that for Al~I
non-LTE leads to higher ionization (overionization) than in LTE in the line formation layers and
to a weakening of spectral lines, which is consistent with earlier non-LTE studies. However, our results, especially for the subordinate lines, differ quantitatively from the results of predecessors. Owing to their large cross sections, the ion-pair production and mutual neutralization processes Al~I($nl$) + H~I($1s$) $\leftrightarrow$ Al~II($3s^2$) + H$^-$ provide a close coupling of high-excitation Al~I levels with the Al~II ground state, which causes smaller deviations from the thermodynamic equilibrium (TE) populations compared to the calculations where
the collisions only with electrons are taken into account. For three mildly metal-poor dwarf
stars, the aluminum abundance has been determined from seven Al~I lines in different models of their
formation. Under the LTE assumption and in non-LTE calculations including the collisions only with
electrons, the Al~I 3961~\AA\ resonance line gives a systematically lower abundance than the mean abundance
from the subordinate lines, by 0.25-0.45~dex. The difference for each star is removed by taking into
account the collisions with hydrogen atoms, and the rms error of the abundance derived from all seven
Al~I lines decreases by a factor of 1.5-3 compared to the LTE analysis. We have calculated the non-LTE abundance corrections for six subordinate Al~I lines as a function of the effective temperature
(4500~K $\le \Teff \le$ 6500~K), surface gravity (3.0 $\le$ log~g $\le$ 4.5), and metal abundance ([M/H] = 0, $-1$, $-2$, and $-3$). For Si~I including the collisions with H~I leads to the establishment of TE populations in the line formation layers even in hot metal-poor models and to minor departures from LTE in spectral lines. 



%

\section{Introduction}

The accuracy of the theoreical profiles and equivalent widths of spectral lines based on the 
non-local thermodynamic equilibrium (non-LTE) line formation depends on the completeness of including all processes of the interaction of a given atom with the radiation field and the surrounding particles in calculating the statistical equilibrium (SE) and on the accuracy of atomic data: the energy levels, photoionization, photoexcitation, and inelastic collision cross sections.
In the atmospheres of solar-type stars, the electron number density ($N_{\rm e}$) is much lower than the number density of neutral hydrogen atoms (H~I). For example, in the solar atmosphere, $N_{\rm e}/N_{\rm H} \simeq 10^{-4}$ in layers with log~$\tau_{5000} < -0.5$. Therefore, the level excitation and
the ion production can result from the collisions not only with electrons but also with H~I. The
importance of including the latter when calculating the SE of a sodium atom was pointed out long ago
by Gehren (1975). Analysis of the observed stellar spectra has shown that in the atmospheres of stars,
especially metal-poor stars, apart from the collisions with electrons, there must exist an additional
source of thermalization, which can be the collisions with H~I atoms (for a review, see Lambert 1993;
Holweger 1996; Mashonkina 2009).

Accurate data for inelastic collisions with H~I are available in the literature only for a small number
of atoms, and they appeared mostly in recent years. The colliding particle is H~I atom in the \ba{1s}{S}{}{} ground state. Below, it will be referred to as H~I for short. The atom being investigated (target) can be in both ground and excited states. Laboratory measurements were made for only one transition, namely Na~I $3s - 3p$ (Fleck et al. 1991). Detailed quantum-mechanical calculations were performed for the Na~I($nl$)+H~I (Belyaev et al. 1999, 2010), Li~I($nl$)+H~I (Belyaev and Barklem 2003), Mg~I($nl$)+H~I (Belyaev et al. 2012), and He~I($nl$)+H~I (Belyaev 2015) collisions. The Al~I($nl$)+H~I and Si~I($nl$)+H~I collisions were calculated by Belyaev (2013) and Belyaev et al. (2014), respectively, with the application of model approaches within the formalism of the Born-Oppenheimer quantum theory. Before the appearance of accurate data, the rates of inelastic collisions with H~I were
calculated using the approximate formulas derived by Steenbock and Holweger (1984) based on the
theory of Drawin (1968). They are commonly called the Drawinian rates in the literature. This approach
continues to be used for most atoms, because there are no other, more accurate methods. It was established empirically that the Drawinian rates require scaling with a factor of \kH\ whose value is different
for different atoms (for references, see Mashonkina et al. 2011). For example, \kH\ $\le 0.01$ for Na~I and Ba~II, but it varies from 0.1 to 30 for other atoms. Recognizing that the criticism of using the Drawinian rates in non-LTE calculations is justified (see, e.g., Barklem et al. 2011), it should be said that this is just a forced necessity.

Many of the chemical elements in the spectra of F-G-K stars are observed in the lines of their neutral
atoms, which contain only a small fraction in the total element abundance. These include Li~I, Na~I,
Mg~I, Al~I, K~I, Fe~I, etc. For example, for aluminum with the ionization energy $\chi_{\rm ion}$ = 6~eV, $N$(Al~I)/$N$(Al) $< 5 \cdot 10^{-3}$ N(Al~I)/N(Al) everywhere in the solar atmosphere. The population
 of these neutral atoms easily deviates from their TE values when the intensity of ionizing radiation deviates from the Planck function, and this affects the spectral line formation. Indeed, calculations for many atoms, from Li~I to Co~I, showed significant non-LTE effects and their growth with decreasing metal abundance (for a review, see Asplund 2005; Mashonkina 2014). The iron abundance
[Fe/H] = log~$N_{\rm Fe}/N_{\rm H} - \log~(N_{\rm Fe}/N_{\rm H})_\odot$ is commonly used as a metallicity indicator. At [Fe/H] $< -2$, the difference in the element abundances derived under the LTE assumption and by abandoning LTE can reach large values. For example, for a star with an effective temperature $\Teff =$ 4680~K, surface gravity log~g = 1.23, and [Fe/H] = $-3$, LTE leads to an abundance from Al~I lines underestimated by 0.18~dex, from Ca~I, on average, by 0.30~dex, and from Ti~I by 0.33~dex, but to a sodium abundance overestimated by 0.23~dex. The number density of electrons in metal-poor atmospheres is low, and they cannot serve as an efficient source of thermalization. Therefore, the magnitude of the non-LTE effects strongly depends on whether and how the inelastic collisions with H~I atoms are taken into account. For example, Mashonkina (2013) determined the magnesium abundance for the star HD~122563
($\Teff$/log~g/[Fe/H] = 4600/1.60/$-2.56$) under the LTE assumption and in non-LTE calculations using
the quantum-mechanical rate coefficients for Mg~I+H~I collisions from Barklem et al. (2012) or including only the electron collisions. For the Mg~I 4703~\AA\ and 5528~\AA\ lines, the non-LTE abundance correction, i.e., the difference between the non-LTE and LTE abundances $\Delta_{\rm NLTE} = \eps{NLTE} - \eps{LTE}$, was +0.27~dex and +0.16~dex, respectively, for the case of pure electron collisions and decreased down to 0.04~dex and $-0.04$~dex, respectively, when the collisions with H~I were taken into account.

This paper is devoted to analyzing the influence of collisions with H~I atoms on the SE and lines of Al~I and Si I. An accurate determination of the aluminum and silicon abundances for stars in a wide metallicity range is very important for testing the models of nucleosynthesis and evolution of the Al and Si abundances of the Galactic matter. Non-LTE calculations for Al~I were performed previously by Baumueller and Gehren (1996), Andrievsky et al. (2008), and Menzhevitski et al. (2012), and they showed that the departures from LTE  have to be taken into account in analyzing the stellar spectra. The influence of non-LTE effects on the silicon abundance determination for the Sun and metal-poor stars was studied by
Shi et al. (2008, 2009). In all these papers, the collisions with H~I were taken into account following Steenbock and Holweger (1984). For the first time we apply the results of accurate calculations
from Belyaev (2013) and Belyaev et al. (2014) and investigate how this affects the abundance determination from individual Al~I and Si I lines as well as the mean abundance and its error.

We describe the methods for calculating a theoretical spectrum of Al~I and Si~I in Section~\ref{sect:method}. Since the departures from LTE turned out to be small for Si~I, the bulk of the paper is devoted to analyzing the Al~I lines. The results are presented in Section\ref{sect:sun} for the Sun and Section~\ref{sect:stars} for stars. The non-LTE abundance corrections for six subordinate Al~I lines as a function of stellar parameters are presented in Section~\ref{NLTE_corr}. The conclusions and recommendations are formulated in Section~\ref{sect:conclusions}.

\section{THE METHODS AND CODES}\label{sect:method}

\subsection{The Al~I Model Atom}

We use the Al~I model atom from Baumueller and Gehren (1996). It includes 59 Al~I levels with a principal quantum number n $\le 15$. The transition probabilities and photoionization cross sections were taken from the OPACITY project. The electron impact excitation is calculated using the formula of van Regemorter (1962) for allowed transitions, and the effective collision strength is assumed to be $\Upsilon$  = 1 for forbidden ones. In contrast to Baumueller and Gehren (1996), who used the Drawinian rates for collisions with H~I, we apply the rate coefficients calculated within the Born-Oppenheimer (BO) approach. Belyaev (2013) considered six lowest Al~I terms up to ba{4d}{D}{}{} (the excitation energy $E_{\rm exc}$ = 4.82~eV) and all the transitions between them caused by collisions with H~I as well as the ion-pair production from the ground and excited states and mutual neutralization Al~I($nl$)+ H~I $\leftrightarrow$ Al~II($3s^2$~$^1$S) + H$^-$ (charged exchange reactions). 

\begin{figure}  
\includegraphics[width=80mm]{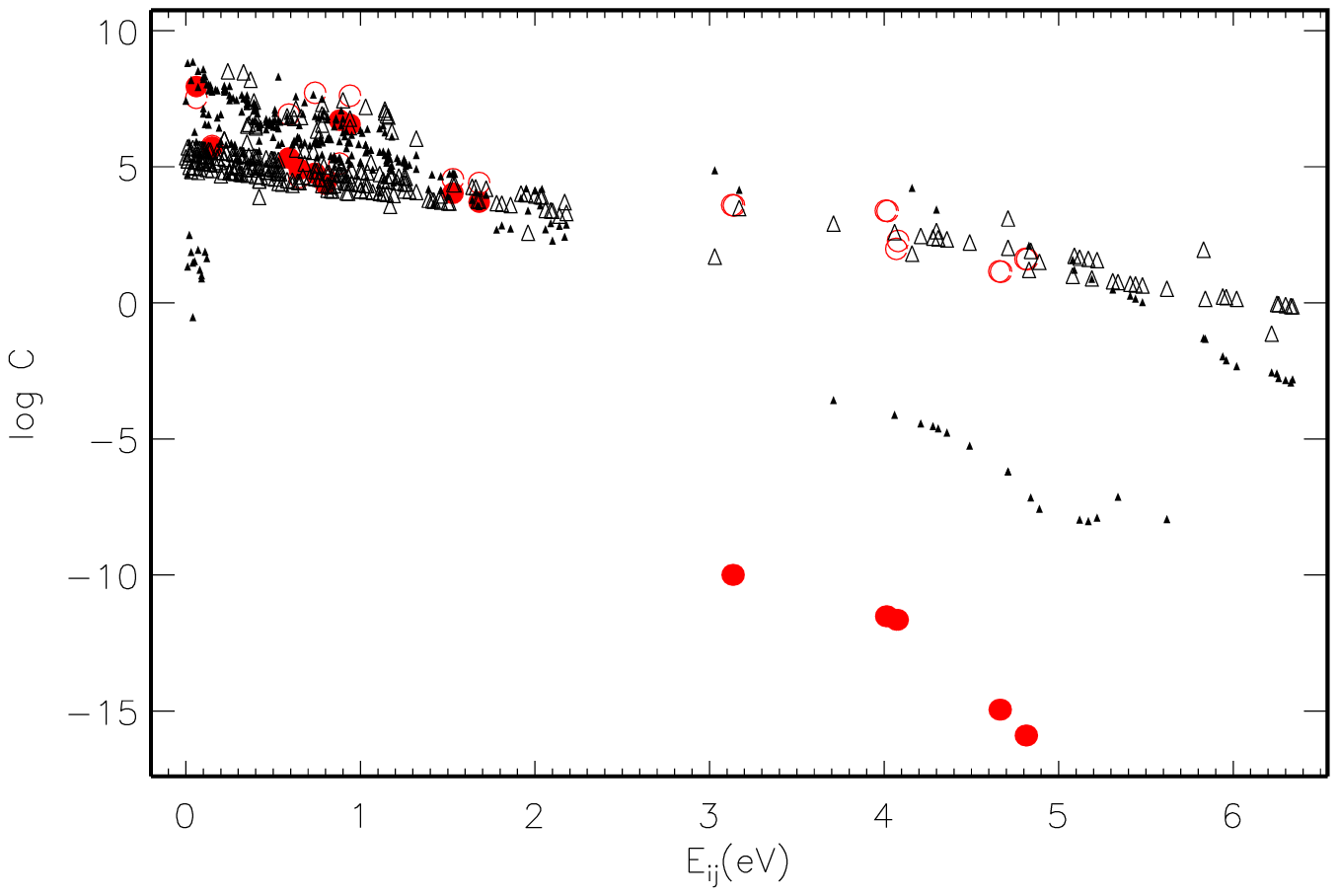}
\includegraphics[width=80mm]{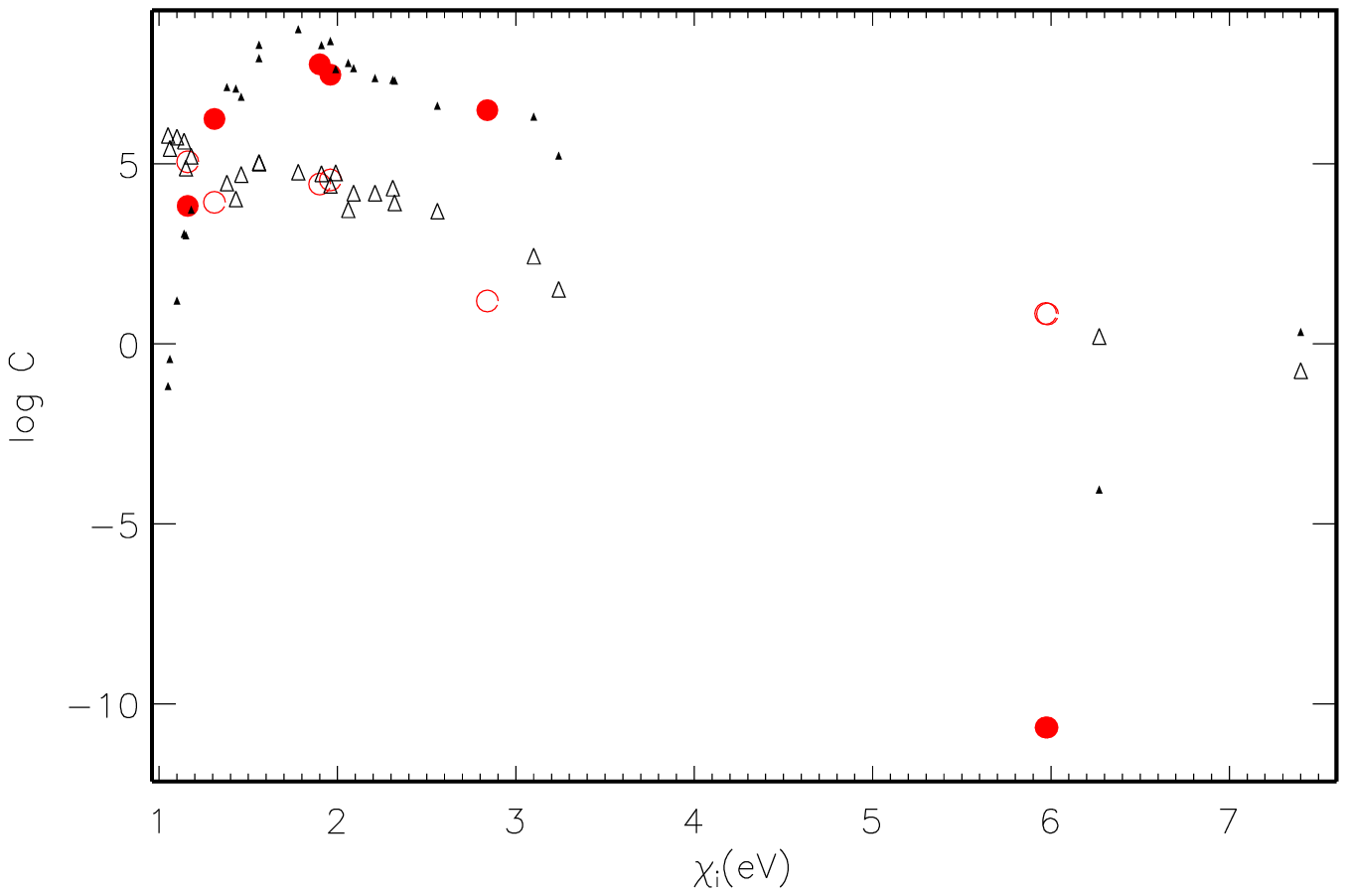}
\caption{Left panel: electron (open symbols) and H~I atom (filled symbols) impact excitation rates for Al~I (circles) and Si~I (triangles). Right panel: rates of the processes $A^0+{\rm e^-} \rightarrow A^++2 e^-$ and $A^0+{\rm H~I} \rightarrow A^+ + {\rm H}^-$ using similar symbols. The calculations were made for $T = 5686$~K, log~$N_{\rm e} = 13$ and log~$N_{\rm H} = 17$.}
\label{fig:rates}
\end{figure}

In Fig.~\ref{fig:rates} we compare the excitation rates due to the collisions with electrons and H~I atoms as a function of the transition energy $E_{\rm ij}$ as well as the ion-pair production rates in the case of collisions with electrons and H~I atoms as a function of the ionization energy of the initial atomic state $\chi_{\rm i}$. The collisions with electrons are more efficient than the collisions
with H~I atoms for bound-bound (b-b) transitions with a large energy separation between the levels 
($E_{\rm ij} > 3$~eV), but the rates $C_{\rm e}$ and $C_{\rm H}$ are comparable in magnitude for $E_{\rm ij} < 2$~eV. The collisions with H~I produce an $A^+ + {\rm H^-}$ ion pair much more efficiently than the collisions with electrons ionize an atom if the atom is in a highly excited state with $\chi_{\rm i} < 3.5$~eV.

\subsection{The Si~I-Si~II Model Atom}

The Si~I-Si~II model atom was constructed by Shi et al. (2008). It includes 132 singlet and triplet Si~I terms, 41 Si~II terms, and the Si~III ground state. Just as for Al~I, the radiative transition probabilities and photoionization cross sections were taken from the OPACITY project. Since there are no accurate data on collisions with electrons, we used the same theoretical approximations as those
for Al~I. To take into account the collisions of neutral
silicon atoms with H~I, we apply the rate coefficients calculated by Belyaev et al. (2014) within the BO approach. For Si~I they considered 25 lower terms up to $4d~^3$F$^\circ$ ($E_{\rm exc}$ = 7.15~eV) and all transitions between them as well as the charge exchange reactions. The excitation and ion-pair production rates due to the collisions with H~I atoms are compared with the electron-impact excitation and ionization rates in Fig.~\ref{fig:rates}. It should be noted that, in contrast to Al~I, there exist many transitions with $E_{\rm ij} > 3$~eV for Si~I, where $C_{\rm H} \simeq C_{\rm e}$. Otherwise, the conclusions for Si~I and Al~I coincide.

Thus, for both Al~I and Si~I a contribution of the inelastic collisions with H~I to establishing the SE is not lesser than that of the collisions with electrons and, consequently, both type collisions must be taken into account in non-LTE calculations.

\begin{figure}  
\includegraphics[width=80mm]{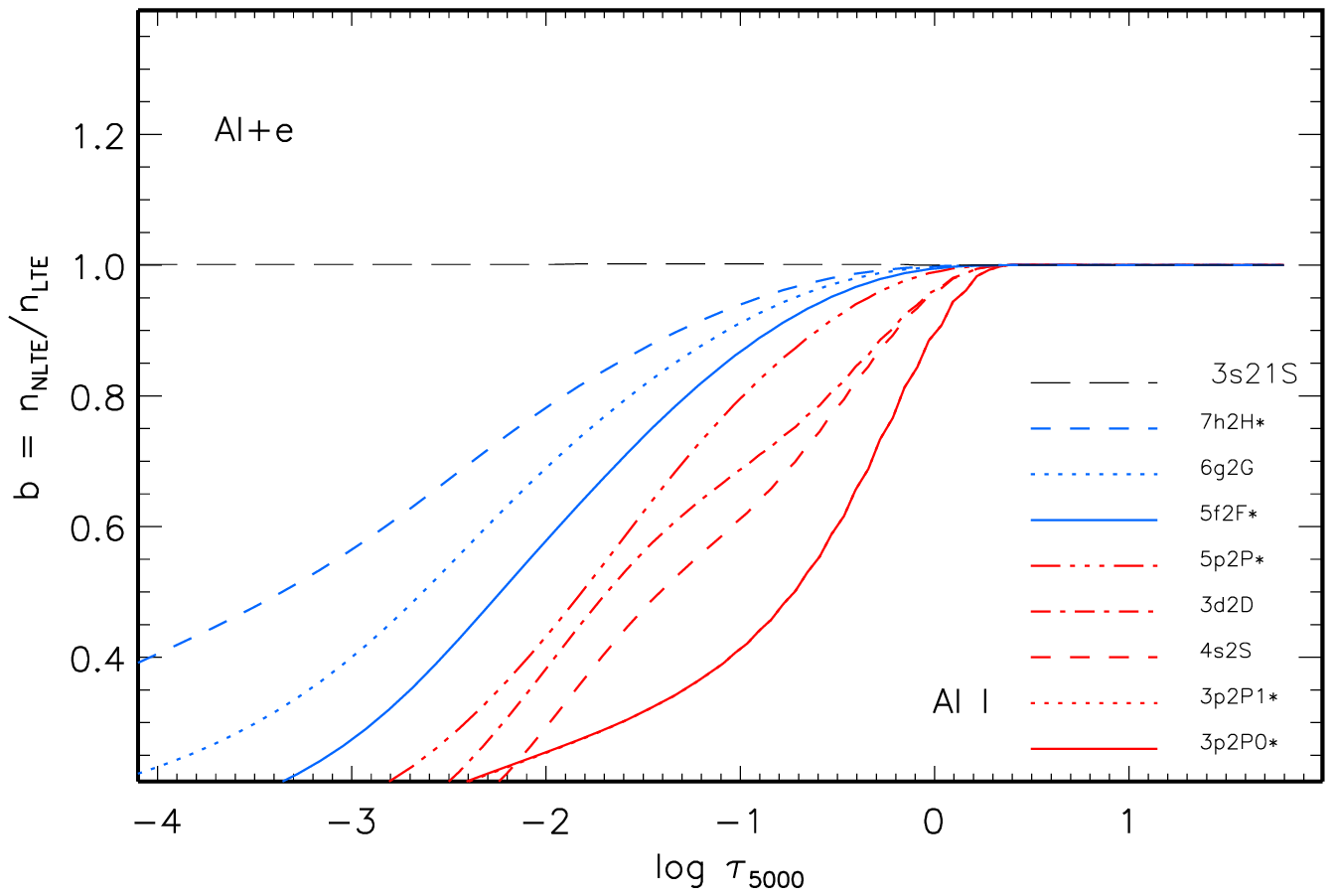}
\includegraphics[width=80mm]{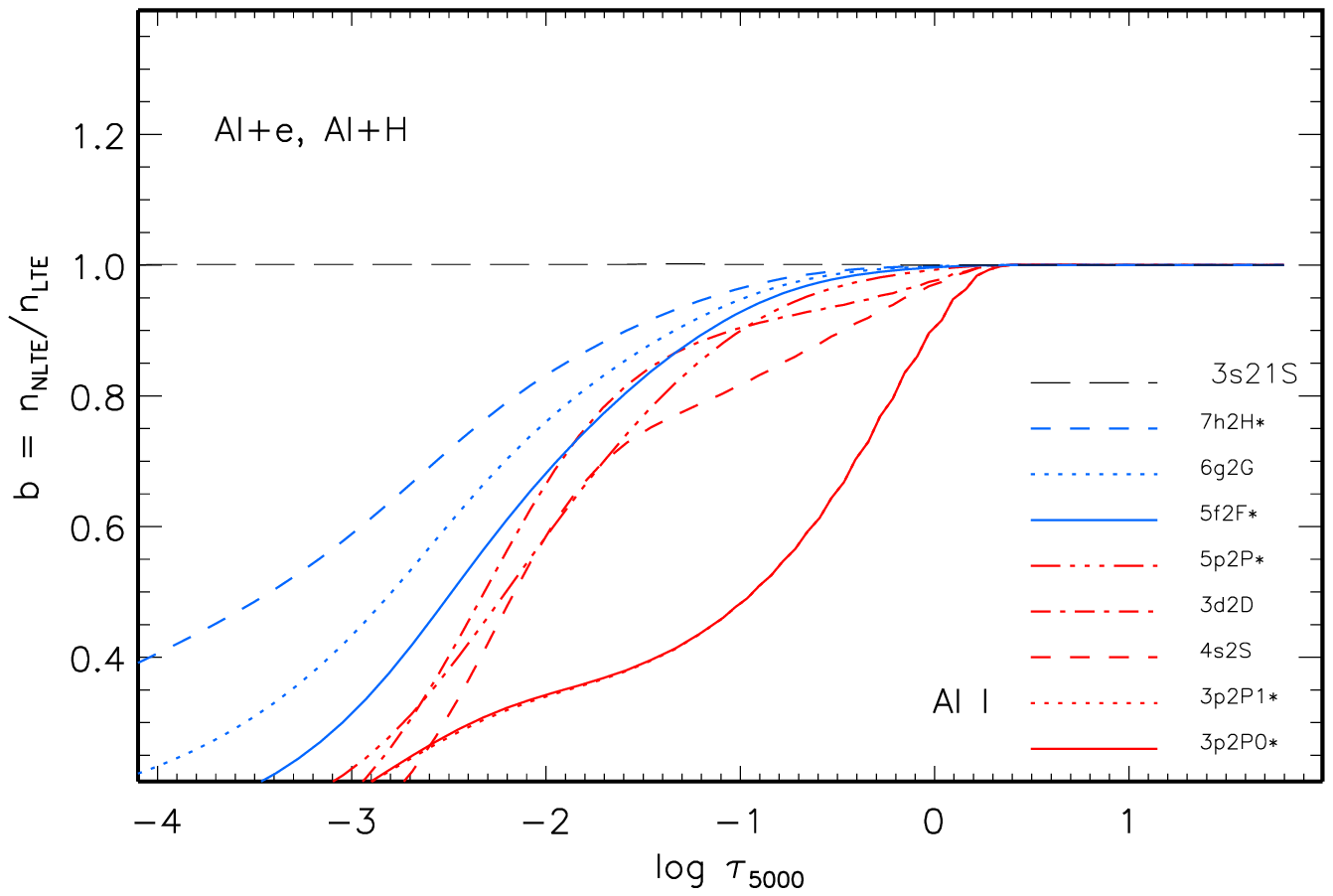}
\caption{The b-factors of selected Al~I levels in the 5890/4.02/$-0.78$ model atmosphere calculated by taking into account the collisions only with electrons (left panel) and including the collisions with H~I atoms (right panel).}
\label{fig:bf}
\end{figure}

\subsection{Non-LTE Calculations}

We solved the system of SE and radiative transfer equations in a specific model atmosphere using
the DETAIL code developed by Butler and Giddings (1985) based on the accelerated $\Lambda$-iteration
method. The level populations obtained by solving the SE equations (non-LTE) and based on the
Boltzmann-Saha formulas (LTE), or rather their ratio b = n$_{\rm NLTE}$/n$_{\rm LTE}$, which we will call below the b-factor, were then used to compute a synthetic spectrum via the SIU (Spectrum Investigation Utility) code (Reetz 1991).

Everywhere in this paper, we used model atmospheres from the MARCS\footnote{\tt http://marcs.astro.uu.se} database (Gustafsson et al. 2008). For given $\Teff$/log~g/[Fe/H], the models were obtained by interpolation using the MARCS algorithm.

Figure~\ref{fig:bf} illustrates the effect of including the collisions with H~I atoms on the statistical equilibrium of Al~I in the 5890/4.02/$-0.78$ model atmosphere. Irrespective of the calculation of collision rates, all Al~I levels are underpopulated relative to the TE values, but the non-LTE effects do not affect the Al~II population. The latter is easy to explain, because Al~II is the dominant ionization stage. As was discussed by Baumueller and Gehren (1996) and shown by other earlier non-LTE studies, the main mechanism of departures from LTE for Al~I is related to an excess of the mean intensity of radiation beyond the ionization threshold for the ground level ($\lambda_{thr} \simeq 2070$~\AA) over the Planck function, which leads to overionization, i.e., excess photoionization compared to the TE one. As expected, including the collisions with H~I atoms leads to a decrease in the departures from LTE for Al~I levels. This is mainly because the coupling of highly excited Al~I levels with the Al~II ground state is strengthened.

\begin{figure}  
\includegraphics[width=80mm]{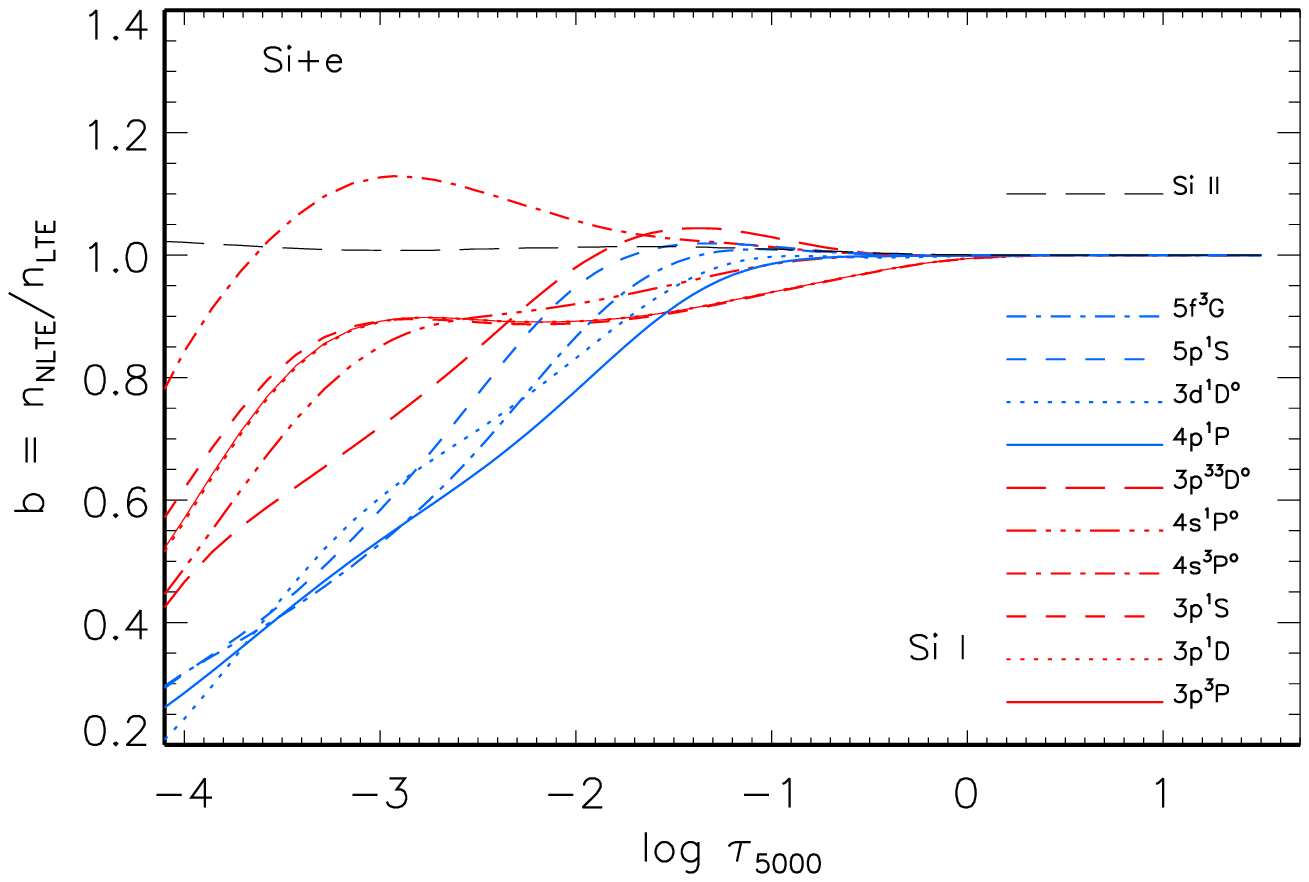}
\includegraphics[width=80mm]{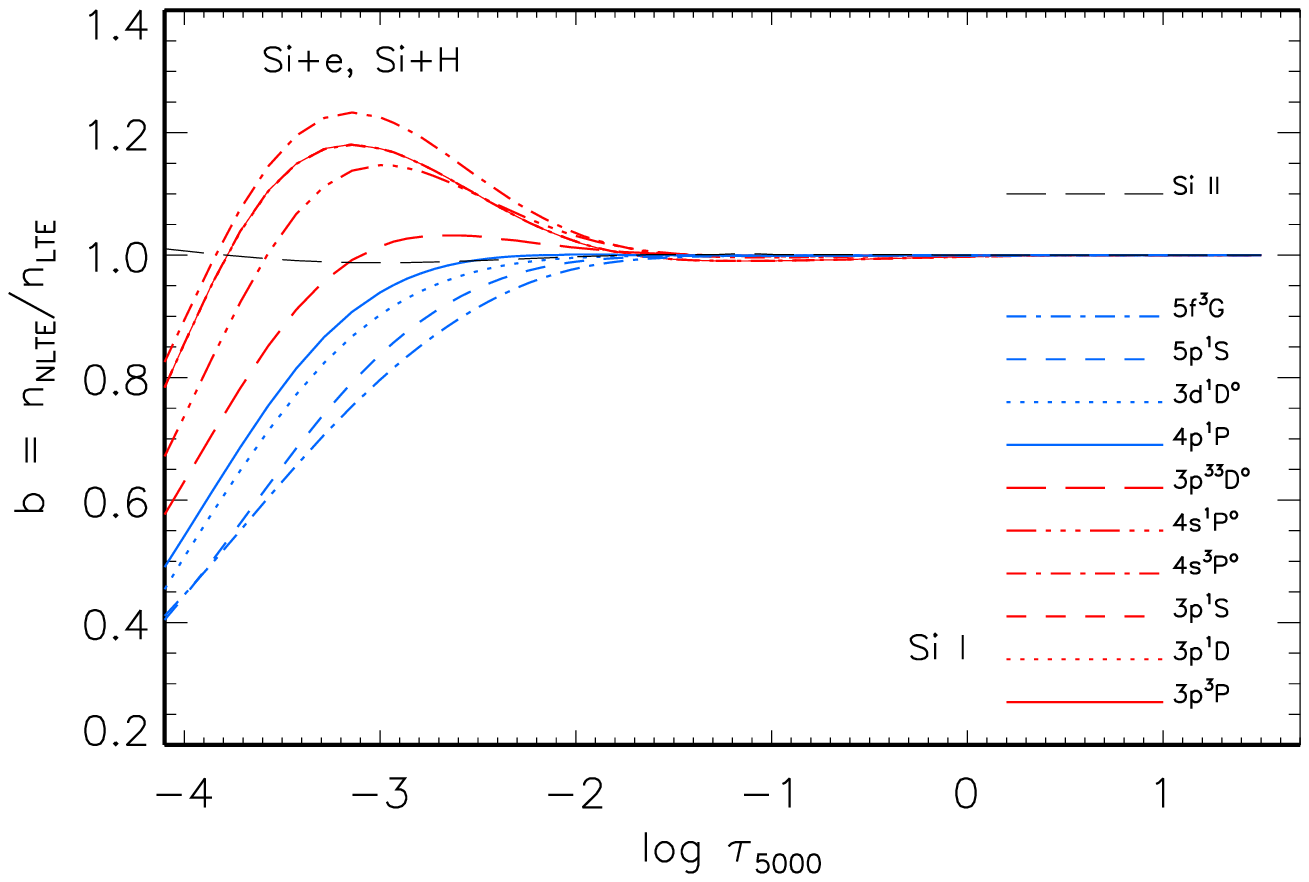}
\caption{Same as Fig.~\ref{fig:bf} for Si~I in the 6115/4.2/$-1.24$ model atmosphere.}
\label{fig:bf_si1}
\end{figure}

For Si~I the ionization energy $\chi$ = 8.19~eV is higher than that for Al~I, and the departures from
LTE are smaller. As an illustration, our non-LTE calculations for Si~I-Si~II were performed with a
higher temperature, $\Teff$ = 6115~K, and a lower metal abundance, [Fe/H] = $-1.24$, than those for Al~I. Figure~\ref{fig:bf_si1} presents the b-factors of selected Si~I levels. We do not show the Si~II levels, because the technique of calculating the collision rates for them did not change. When the collisions only with electrons are taken into account, Si~I shows a typical picture of overionization at line formation depths (llog$\tau_{5000} < 0$). The $4s ^3$P$^\circ$ level in the outermost layers constitutes
an exception. Including the collisions with Рќ~I atoms removes the departures from LTE in layers deeper
than log$\tau_{5000} = -2$. In the uppermost layers, the populations of the ground and low-excitation Si~I states are higher than the TE ones, i.e., there is an effect opposite to overionization.

The departures from LTE for the Si~I lines are small independent of whether the collisions
with Рќ~I atoms are taken into account or are not. For example, in the 6115/4.2/$-1.24$ model,
the non-LTE corrections for the Si~I 4102\,\AA\ ($3p ^1$S$_0$ -- $4s ^3$P$^{\circ}_{1}$), 3905\,\AA\ ($3p ^1$S$_0$ -- $4s ^1$P$^{\circ}_{1}$), 5772\,\AA\ ($4s ^1$P$^{\circ}_{1}$ -- $5p ^1$S$_0$), and 6155\,\AA\ ($3p^3\, ^3$D$^{\circ}_{3}$ -- $5f ^3$G$_4$) lines are $\Delta_{\rm NLTE}$ = 0.12, 0.03, 0.00, and $-0.03$~dex, respectively, in the case of pure electronic collisions and become vanishingly small when the collisions with Рќ~I atoms are included: $\Delta_{\rm NLTE}$ = 0.00, 0.00, $-0.01$, and $-0.02$~dex.

Therefore, our subsequent discussion will be devoted to analyzing the departures from LTE in Al~I lines and the role of accurate data on the Al~I + Рќ~I collisions in improving the aluminum abundance in stellar atmospheres.

\section{ANALYSIS OF Al~I LINES IN THE SOLAR SPECTRUM}\label{sect:sun}

\begin{figure}  
\includegraphics[width=100mm]{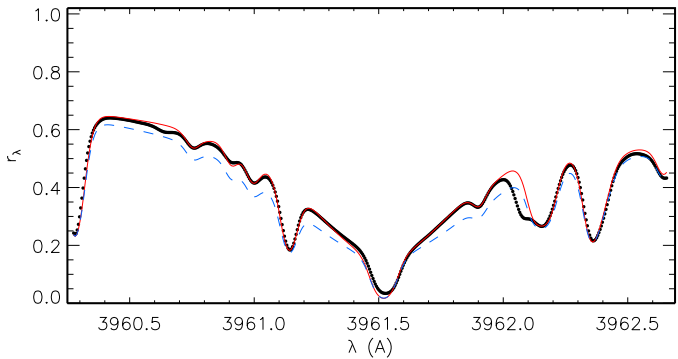}
\caption{The Al~I 3961~\AA\ line profile in the solar spectrum (Kurucz 2005; filled circles) in comparison with the theoretical non-LTE profiles computed by taking into account the collisions with H~I atoms (the BO case) at different log~$C_6 = -31.2$ (solid curve) and $-30.62$ (dashed curve). The aluminum abundance in the calculations is $\eps{Al}$ = 6.45, the model atmosphere is 5780/4.44/0.0, and $\xi_{\rm t}$ = 0.9~\kms.}
\label{fig:sun3961}
\end{figure}

We analyze the spectrum of the Sun as a star (Kurucz 2005) and use the abundances derived from individual lines to determine the differential abundances for stars. All our calculations were made with
the 5780/4.44/0.00 model atmosphere and microturbulence $\xi_{\rm t}$ = 0.9~\kms. Our analysis is based on fitting the observed line profiles rather than equivalent widths. An example of the Al~I 3961~\AA\ line is shown in Fig.~\ref{fig:sun3961}. For comparison with observations, the theoretical profile was convolved with the rotation profile and the radial-tangential profile of macroturbulent motions. The rotation velocity was taken to be $v_{rot} = 1.8$~\kms, and the most probable velocity of the radial and tangential motions is $\Vmac$ = 3.5~\kms.

In F-G-K stars, neutral aluminum is observed in the Al~I 3944, 3961~\AA\ resonance doublet and
subordinate, weaker lines in the red and infrared (IR) spectral ranges. Their list together with the atomic parameters and data sources is given in Table~\ref{tab_sun}. The Al~I 3944~\AA\ line is blended with a strong РЎРќ molecular line and is not used to determine the abundance. Note that with the van der Waals broadening constant log~$C_6 = -30.62$ calculated for Al~I 3961~\AA\ based on the data from Anstee and O'Mara (1995), the solar non-LTE abundance from this line, $\eps{Al}$ = 6.26, is too low compared to the meteoritic $\eps{Al}$ = 6.45 (Lodders et al. 2009). It is even lower, by 0.06~dex, in LTE. Here, we use the scale where $\eps{H}$ = 12 for hydrogen. Taking $\eps{Al}$ = 6.45 as the solar abundance, we found the observed Al~I 3961~\AA\ profile to be reproduced in the non-LTE calculations if log~$C_6 = -31.2$ (Fig.~\ref{fig:sun3961}). This value will be used in further analyzing the solar and stellar spectra.

\begin{table}[htbp]
\caption{Atomic parameters of the investigated Al~I lines and the solar abundance derived in different line formation scenarios.}
\label{tab_sun}
\tabcolsep3.7mm
\begin{center}
\begin{tabular}{ccccccccc}
\hline\noalign{\smallskip}
$\lambda$ & E$_{exc}$ & $\log gf$ & Ref. & $\log C_6$  & Ref. & \multicolumn{3}{c}{Solar abundance, $\eps{Al}$ } \\ 
\cline{7-9}   
(\AA ) & (eV) &   &  &  &  & LTE & (BO)$^1$ & (e)$^2$ \\
\hline\noalign{\smallskip}
3961.52 & 0.01 &  $-$0.33 & NIST & $-$31.2 & S   & 6.39 & 6.45 & 6.45 \\
6696.02 & 3.14 &  $-$1.35 & WSM & $-$30.52 & K12 & 6.27 & 6.29 &  6.32 \\
6698.67 & 3.14 &  $-$1.65 & WSM & $-$30.52 & K12 & 6.25 & 6.27 &  6.30 \\
7835.31 & 4.02 &  $-$0.61 & K12 & $-$29.77 & K12 & 6.38 & 6.40 &  6.44 \\
7836.13 & 4.02 &  $-$0.46 & K12 & $-$29.77 & K12 & 6.45 & 6.47 &  6.51 \\ 
8772.86 & 4.02 &  $-$0.21 & K12 & $-$29.01 & B98 & 6.31 & 6.32 &  6.37 \\
8773.90 & 4.02 &  $-$0.06 & K12 & $-$29.01 & B98 & 6.37 & 6.38 &  6.42  \\
\noalign{\smallskip}\hline \noalign{\smallskip}
\multicolumn{9}{l}{ \ $^1$ Non-LTE, the collisions with H~I are taken into account according to Belyaev (2013). } \\
\multicolumn{9}{l}{ \ $^2$ Non-LTE, the collisions only with electrons are taken into account.} \\
\multicolumn{9}{l}{Ref.: NIST = Ralchenko et al. (2008); WSM = Wiese et al. (1969);} \\
\multicolumn{9}{l}{S = solar line profile fitting, B98 = Barklem et al. (1998);} \\
\multicolumn{9}{l}{K12 = {\tt http://kurucz.harvard.edu/atoms.html}} \\
\end{tabular}
\end{center}
\end{table}  %

We found the mean abundance from six subordinate lines to be $\eps{Al}$ = 6.36$\pm$0.07 in the non-LTE
calculations (BO), 6.39$\pm$0.08 in the non-LTE calculations with pure electron collisions (case e), and 6.34$\pm$0.07 in LTE. The rms deviation $\sigma_{\varepsilon} = \sqrt{\Sigma(\overline{x}-x_i)^2/(N-1)}$ is given everywhere as the statistical error. The abundance derived under the LTE assumption is lower than the meteoritic one by 0.11~dex. The non-LTE calculations lead to better agreement between the photospheric and meteoritic abundances, though it should be noted that the departures from LTE for Al~I lines in the solar atmosphere are small.

\section{THE ALUMINUM ABUNDANCE IN SELECTED STARS}\label{sect:stars}

From an observational point of view, the aluminum abundance is difficult to determine in F-G-K stars,
especially if the sample includes stars in a wide metallicity range. The Al~I 3961.52~\AA\ resonance line is located in the Ca~II K 3968.47~\AA\ line wing and is difficult to measure in close-to-solar metallicity or mildly metal-deficient stars, because the Ca~II~K line is very strong, and the problems of
normalizing the observed spectrum arise. In contrast, the subordinate Al~I lines become unmeasurable in
stars with [Fe/H] $< -1$. This means that for a stellar sample covering, say, the range $-3 <$ [Fe/H] $< 0$, the aluminum abundance is determined from different lines at the opposite ends of the metallicity scale, and we should be sure that using different lines does not lead to a systematic shift in the abundance. In this paper, we choose three mildly metal-poor stars for which high-resolution spectra are available in the entire spectral range where the Al~I lines are located and investigate how the application of accurate collisional data affects the abundance derived from individual lines and the agreement between the results from different lines.

\begin{figure}  
\includegraphics[width=80mm]{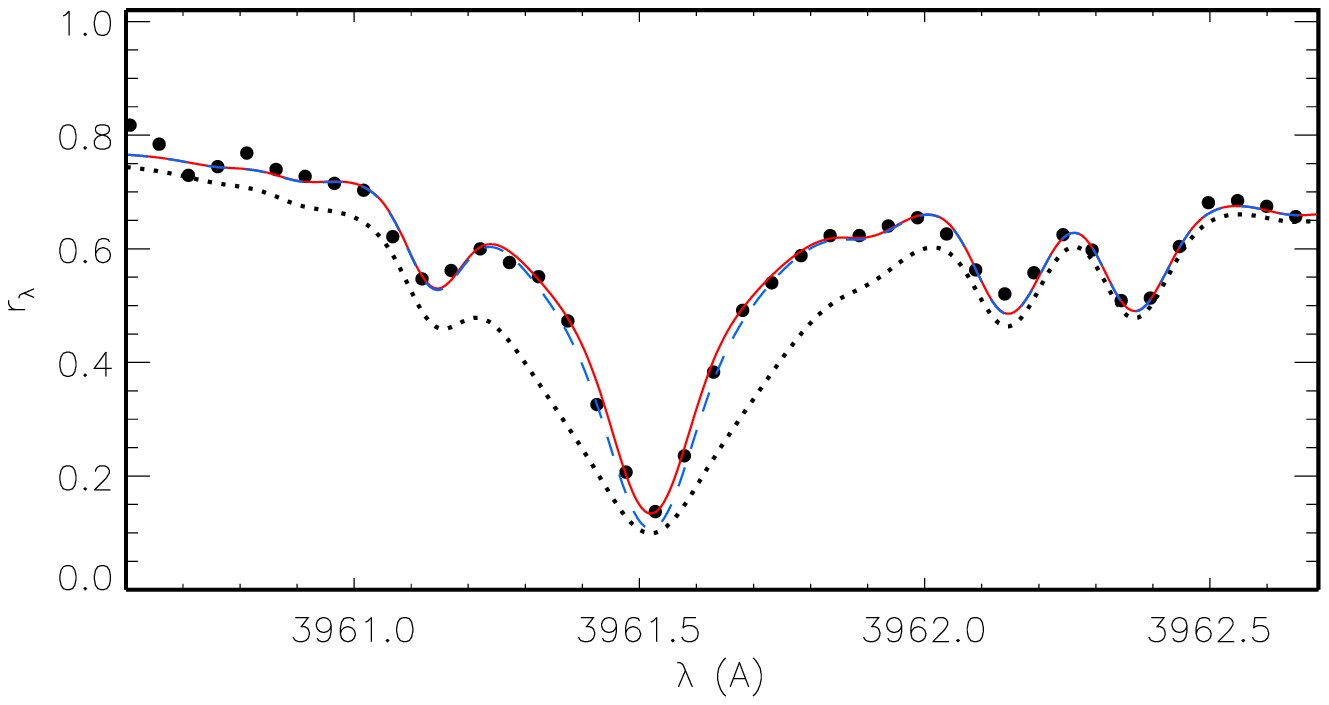}
\includegraphics[width=80mm]{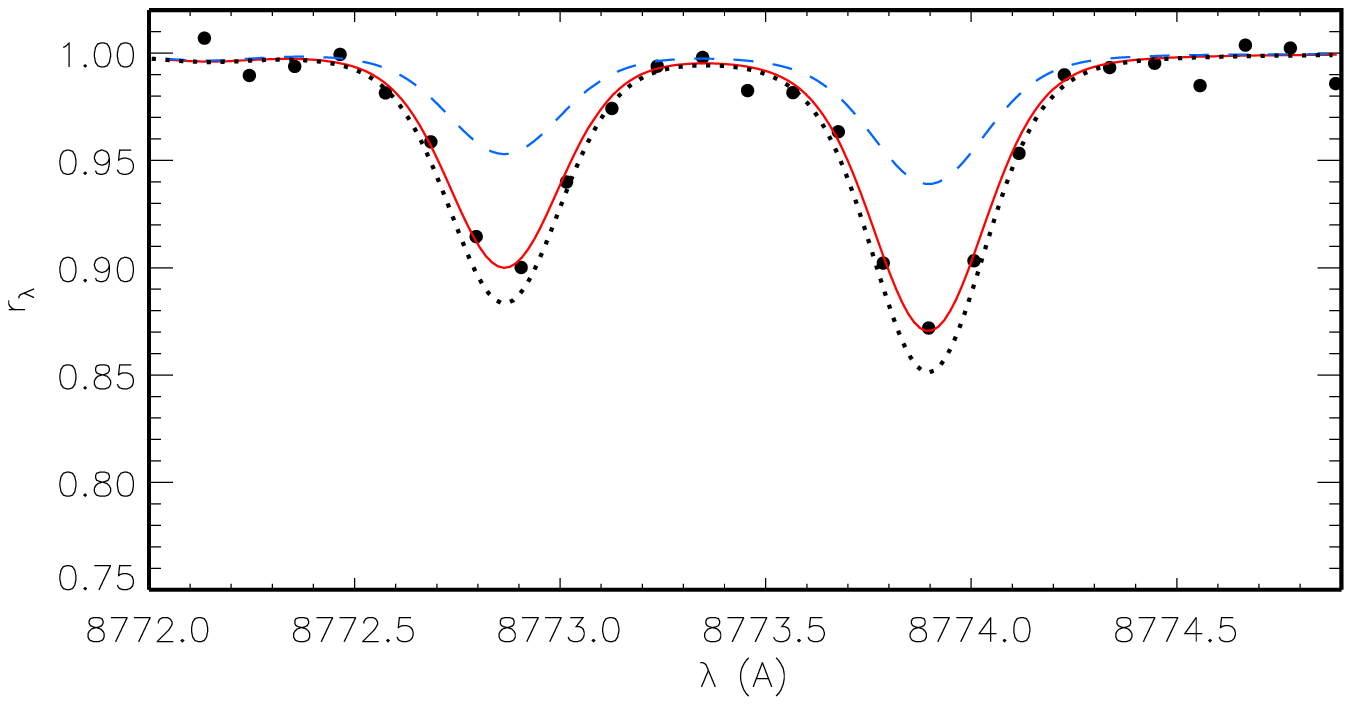}
\caption{The Al~I line profiles in the spectrum of HD~134169 (filled circles) in comparison with the theoretical ones calculated with the 5890/4.02/$-0.78$ model atmosphere in different line formation scenarios: LTE (dotted curve), non-LTE(BO) (solid curve), and non-LTE(e) (dashed curve). The aluminum abundance in the calculations: [Al/H] = $-0.34$ for Al~I 3961~\AA\ (left panel) and [Al/H] = $-0.44$ for Al~I 8772, 8773~\AA\ (right panel).}
\label{fig:profiles}
\end{figure}

\begin{table}[htbp]
\caption{Stellar aluminum abundances [Al/H] in different models of line formation for Al~I.}
\label{tab_stars}
\tabcolsep3.7mm
\begin{center}
\begin{tabular}{rrrrrrrrc}
\hline\hline\noalign{\smallskip}
 & \multicolumn{7}{c}{Al~I lines } & Mean \\
 & 3961 & 6696 & 6698 & 7835 & 7836 & 8772 & 8773 & \\ 
\cline{2-8}   
Model & \multicolumn{8}{l}{HD~59374: 5850/4.38/$-0.88$, $\xi_{\rm t}$ = 1.2~\kms} \\  
\cline{2-9}   
LTE  & $-$0.97 & $-$0.75 & $-$0.79 & $-$0.66 & $-$0.76 & $-$0.60 & $-$0.63 & $-$0.74$\pm$0.12 \\
(BO)$^1$ & $-$0.48 & $-$0.64 & $-$0.70 & $-$0.60 & $-$0.70 & $-$0.54 & $-$0.57 & $-$0.60$\pm$0.08 \\
(e)$^2$ & $-$0.55 & $-$0.42 & $-$0.48 & $-$0.21 & $-$0.27 & $-$0.11 & $-$0.08 & $-$0.30$\pm$0.18 \\
(Dr*0.1)$^3$ & $-$0.55 & $-$0.56 & $-$0.61 & $-$0.51 & $-$0.61 & $-$0.42 & $-$0.44 & $-$0.53$\pm$0.08 \\
\cline{2-9}   
Model & \multicolumn{8}{l}{HD~59984: 5930/4.02/$-0.69$, $\xi_{\rm t}$ = 1.4~\kms} \\ \cline{2-9} 
LTE  & $-$1.05 &   &   & $-$0.58 & $-$0.71 & $-$0.60 & $-$0.63 & $-$0.71$\pm$0.19 \\
(BO) & $-$0.61 &   &   & $-$0.50 & $-$0.63 & $-$0.51 & $-$0.54 & $-$0.56$\pm$0.06 \\
(e)  & $-$0.69 &   &   & $-$0.20 & $-$0.30 & $-$0.15 & $-$0.13 & $-$0.29$\pm$0.23 \\
(Dr*0.1) & $-$0.66 & & & $-$0.43 & $-$0.54 & $-$0.38 & $-$0.39 & $-$0.48$\pm$0.12 \\
\cline{2-9}   
Model & \multicolumn{8}{l}{HD~134169: 5890/4.02/$-0.78$, $\xi_{\rm t}$ = 1.2~\kms} \\ \cline{2-9}
LTE  &  $-$1.11 & $-$0.72 & $-$0.80 & $-$0.62 & $-$0.65 & $-$0.50 & $-$0.53 & $-$0.70$\pm$0.21 \\
(BO) &  $-$0.34 & $-$0.60 & $-$0.67 & $-$0.54 & $-$0.56 & $-$0.41 & $-$0.44 & $-$0.51$\pm$0.12 \\
(BObf)$^4$ &  $-$0.42 & $-$0.58 & $-$0.65 & $-$0.52 & $-$0.54 & $-$0.38 & $-$0.41 & $-$0.50$\pm$0.10 \\
(e)  &  $-$0.43 & $-$0.39 & $-$0.47 & $-$0.14 & $-$0.10 &  0.04 &  0.10 & $-$0.20$\pm$0.23 \\
(Dr*0.1) & $-$0.41 & $-$0.47 & $-$0.55 & $-$0.45 & $-$0.45 & $-$0.25 & $-$0.25 & $-$0.40$\pm$0.11 \\
\noalign{\smallskip}\hline \noalign{\smallskip}
\multicolumn{9}{l}{ \ $^1$ Non-LTE, the Al + H collisions according to Belyaev (2013), } \\
\multicolumn{9}{l}{ \ $^2$ Non-LTE, the collisions only with electrons.} \\
\multicolumn{9}{l}{ \ $^3$ Non-LTE, the Al + H collisions with the Drawinian rates, \kH\ = 0.1,} \\
\multicolumn{9}{l}{ \ $^4$ Non-LTE, only ${\rm Al~I + H~I \leftrightarrow Al^+ + H^-}$. } \\
\end{tabular}
\end{center}
\end{table} 

The stellar spectra were taken by T. Gehren at the Calar Alto Observatory (Spain) using the 2.2-m telescope with the FOCES spectrograph. The spectral range is 3800-8800~\AA, the spectral resolution
is $R = \lambda/\Delta\lambda \simeq$ 40 000, and the signal-to-noise
ratio is $S/N \simeq 200$. The IR spectra were reduced by H. Zhang.

The atmospheric parameters were taken from Sitnova et al. (2015). The effective temperatures
and surface gravities were determined using several methods, namely by the IR flux method ($\Teff$), by
invoking the trigonometric parallaxes (log~g), and spectroscopically through a non-LTE analysis of Fe~I
and Fe~II lines ($\Teff$, log~g). The iron abundance and microturbulence were determined simultaneously
with $\Teff$ and log~g in the non-LTE analysis of Fe~I and Fe~II lines. The data are given in Table~\ref{tab_stars}.

The aluminum abundance in the stars was determined under the LTE assumption and in our non-LTE calculations by taking into account the collisions only with electrons (e) and including the collisions
with H~I atoms with accurate data from Belyaev (2013, BO). For each line, we applied a differential
approach, i.e., the abundance was determined relative to the solar one: [Al/H] = log~$gf\varepsilon_{Р·РІ}$ -- log~$gf\varepsilon_{\odot}$. The results are presented in Table~\ref{tab_stars}. Just as for the Sun, the stellar spectra were analyzed by the synthetic-spectrum method. The theoretical profile
was convolved with the instrumental profile and the radial-tangential microturbulence profile. For
the instrumental profile, we used a Gaussian with a
dispersion of 4~\kms; $\Vmac$ was varied within 10~\%\ near 4.5~\kms\ for the subordinate lines and was
larger, 6 and 8~\kms\ for the resonance line. Since all stars rotate slowly ($v_{rot} \sin~i \le 1$~\kms), the rotational broadening was disregarded. An example of the observed spectra and their best non-LTE fits is shown in Fig.~\ref{fig:profiles}.

\begin{figure}  
\resizebox{80mm}{!}{\includegraphics{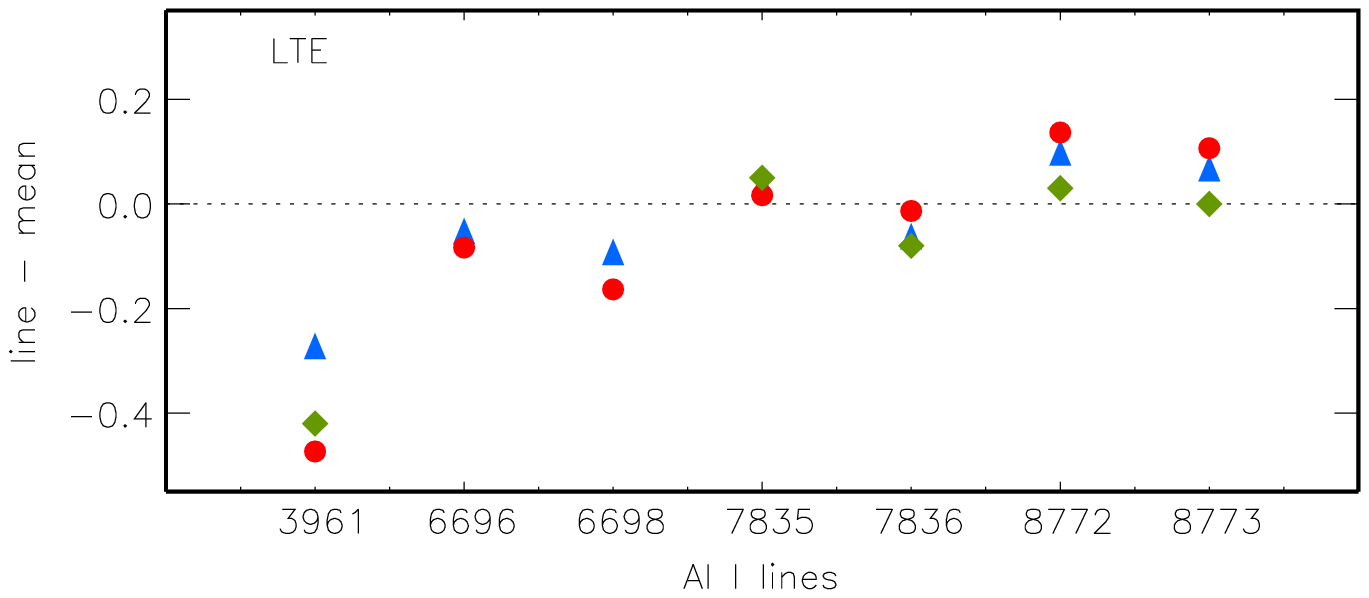}}
\resizebox{80mm}{!}{\includegraphics{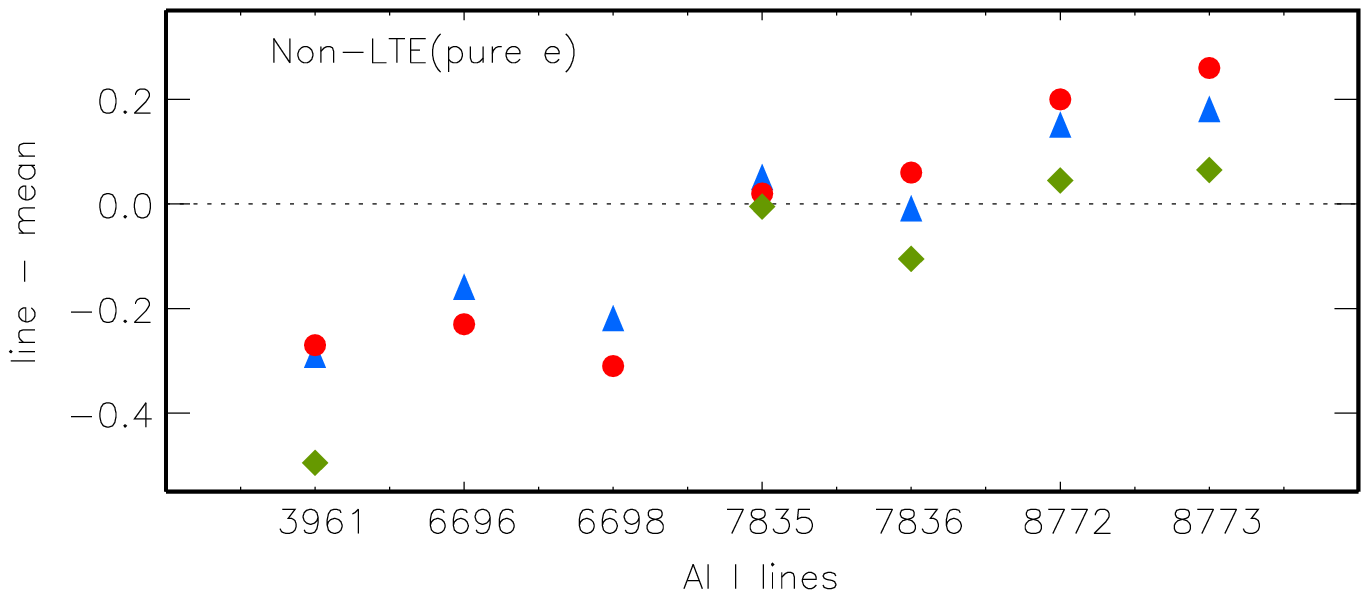}}
\resizebox{80mm}{!}{\includegraphics{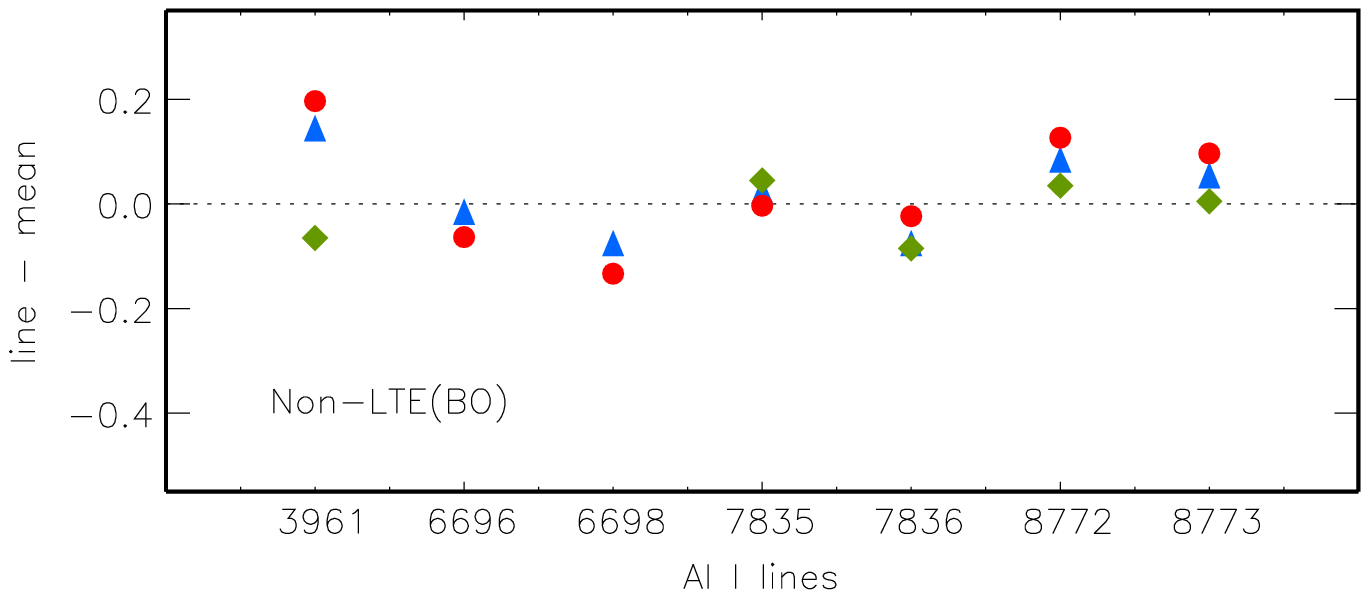}}
\caption{Difference in abundance between each Al~I line and the mean value obtained from the subordinate lines in different line formation scenarios: LTE (left top panel), non-LTE including the collisions only with electrons (right top panel), and non-LTE including the collisions with H~I atoms (bottom panel). Different symbols correspond to the stars HD~59374 (triangles), HD~59984 (diamonds), and HD~134169 (circles).}
\label{fig:diff7836}
\end{figure}

Let us analyze how the abundances from different lines agree in different Al~I line formation models.
The LTE calculations yield a large scatter in abundances between different lines with $\sigma_{\varepsilon}$ from 0.12 to 0.21~dex. This is also illustrated by Fig.~\ref{fig:diff7836}, where the difference in abundance between each line and the mean value from the subordinate lines is presented.
The subordinate lines give a systematically higher LTE abundance than does Al~I 3961~\AA, by 0.25-0.45~dex for different stars. This means that systematic shifts will arise when the aluminum
abundance is determined for a sample of stars under the LTE assumption if we use the subordinate lines
for stars with [Fe/H] $> -1$ but Al~I 3961\AA\ for more metal-poor stars.

Non-LTE leads to a dramatic weakening of the resonance line in all models and a positive non-LTE abundance correction with $\Delta_{\rm NLTE}$ from 0.36 to 0.77~dex. In non-LTE(e), the non-LTE corrections for the subordinate lines are almost equally large and the scatter in the abundances from different lines turns out to be even greater than that in LTE, so that $\sigma_{\varepsilon}$ reaches 0.18-0.23 dex. Including the collisions with H~I leads to a decrease in the departures from LTE and the non-LTE corrections for the subordinate lines. For example, for Al~I 8773~\AA\ in the star HD~134169, $\Delta_{\rm NLTE}$ = 0.63~dex in non-LTE(e) and 0.09~dex in non-LTE(BO). For Al~I 6696~\AA\ in the same star, the corresponding values are 0.33 and 0.12~dex. In non-LTE(BO), the mean abundance for different stars increases compared to LTE by 0.14-0.19~dex, the scatter in the abundances from different lines decreases, and the error becomes smaller by a factor of 1.5-3. These calculations prove conclusively that inelastic collisions with H~I affect the SE of aluminum and must be taken into account when calculating a theoretical spectrum. The application of accurate collisional data allows the astrophysical results to be improved.

We performed test calculations by excluding the excitation processes due to the collisions with H~I.
The results for HD~134169 are given in Table~\ref{tab_stars} (BObf). The departures from LTE for the subordinate lines increased, so that $\Delta_{\rm NLTE}$ increased by 0.02-0.03~dex. In contrast, for the
resonance line, $\Delta_{\rm NLTE}$ decreased by 0.08~dex, and the abundance turned out to be the same as in the case of pure electronic collisions. These results suggest that all processes of the interaction
between Al~I and H~I atoms need to be taken into account.

Since the collisions with H~I for most atoms are still taken into account using the formulas from
Steenbock and Holweger (1984), we also determined the aluminum
abundance in our non-LTE calculations with scaled (\kH\ = 0.1) Drawinian rates, (Dr*0.1) in Table~\ref{tab_stars}. For the resonance line, the non-LTE effects turned out to be very close to the case where the collisions only with electrons are taken into account. For the subordinate lines, the departures from LTE decreased compared to non-LTE(e) but increased compared to non-LTE(BO). The mean abundance in non-LTE(e) turns out to be larger than that in non-LTE(BO) by 0.27-0.31~dex, while the abundance difference (Dr*0.1 - BO) is between 0.07 and 0.11~dex for different stars. Thus, it is better to use the approximate formulas than to disregard the collisions with H~I.

\section{THE NON-LTE CORRECTIONS FOR Al~I LINES AS A FUNCTION OF STELLAR
PARAMETERS}\label{NLTE_corr}

\begin{figure}  
\resizebox{50mm}{!}{\includegraphics{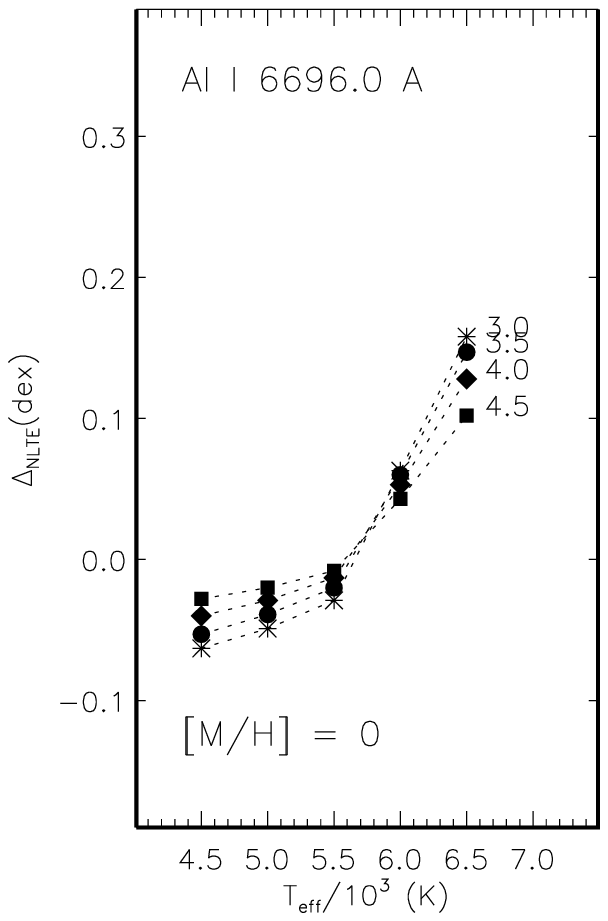}}
\resizebox{50mm}{!}{\includegraphics{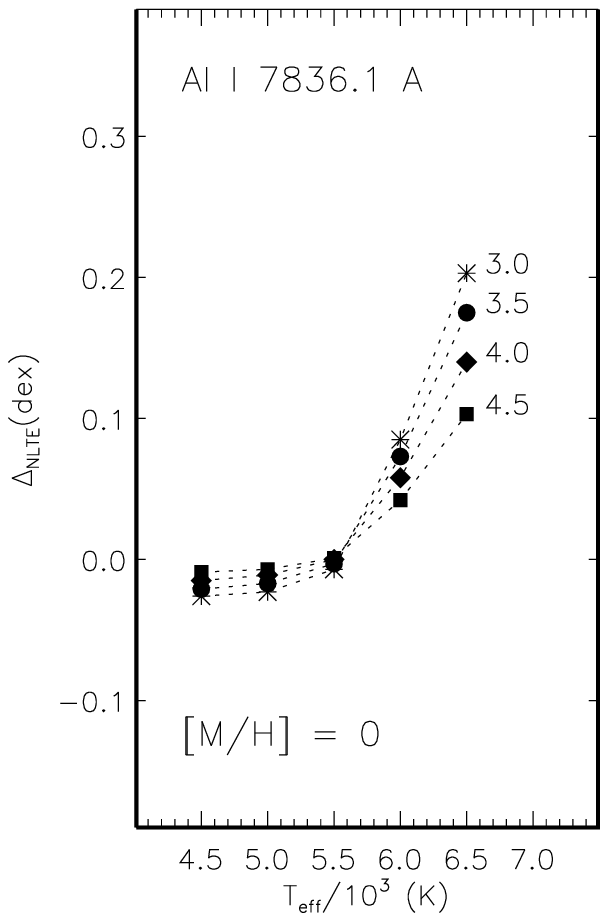}}
\resizebox{50mm}{!}{\includegraphics{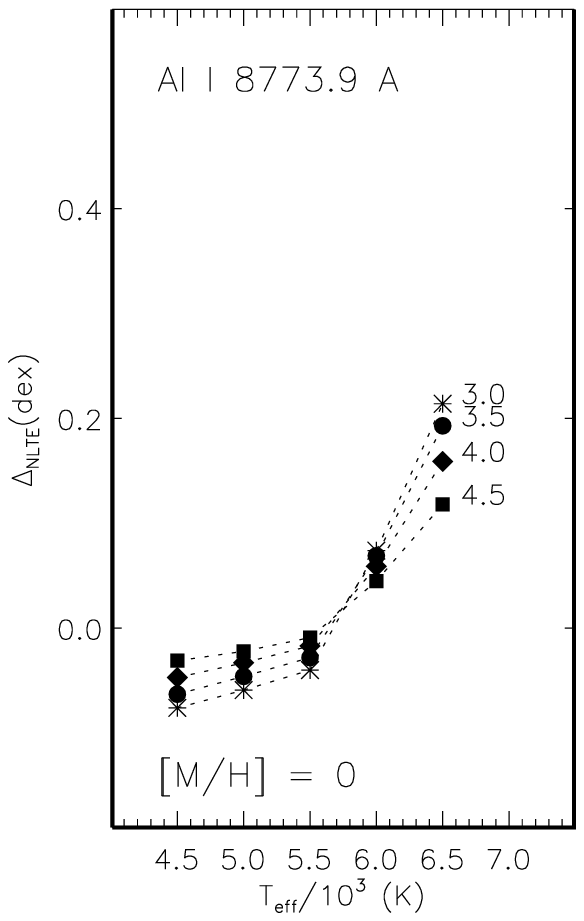}}
\resizebox{50mm}{!}{\includegraphics{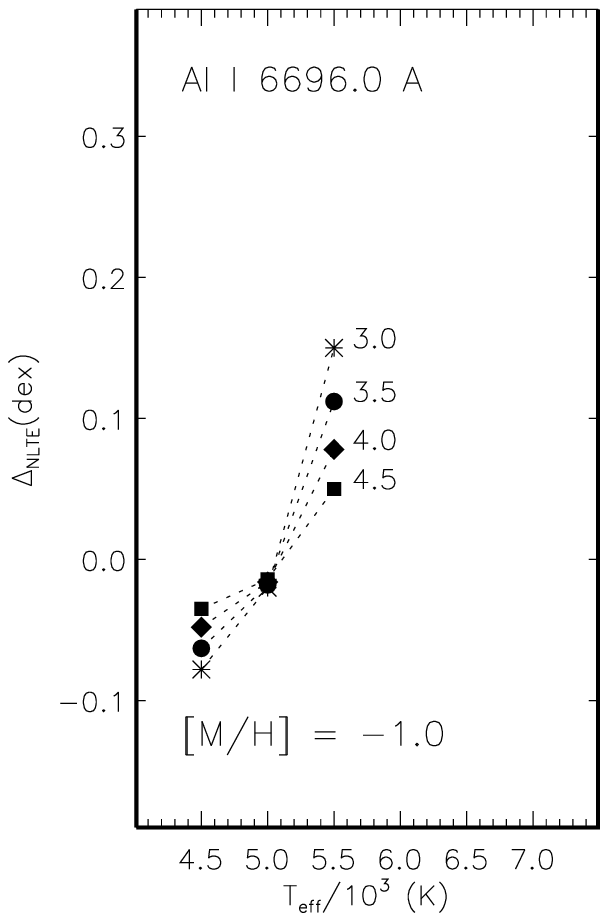}}
\resizebox{50mm}{!}{\includegraphics{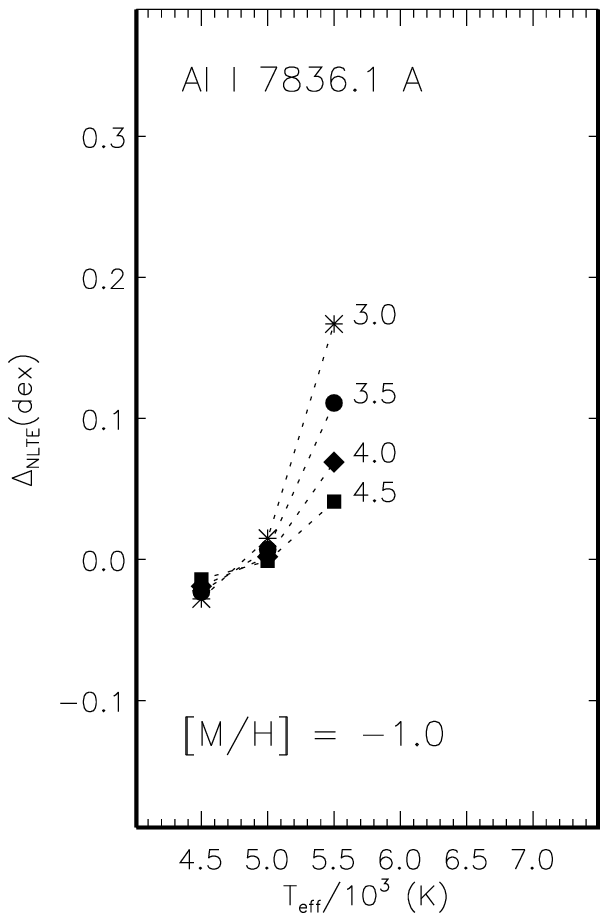}}
\resizebox{50mm}{!}{\includegraphics{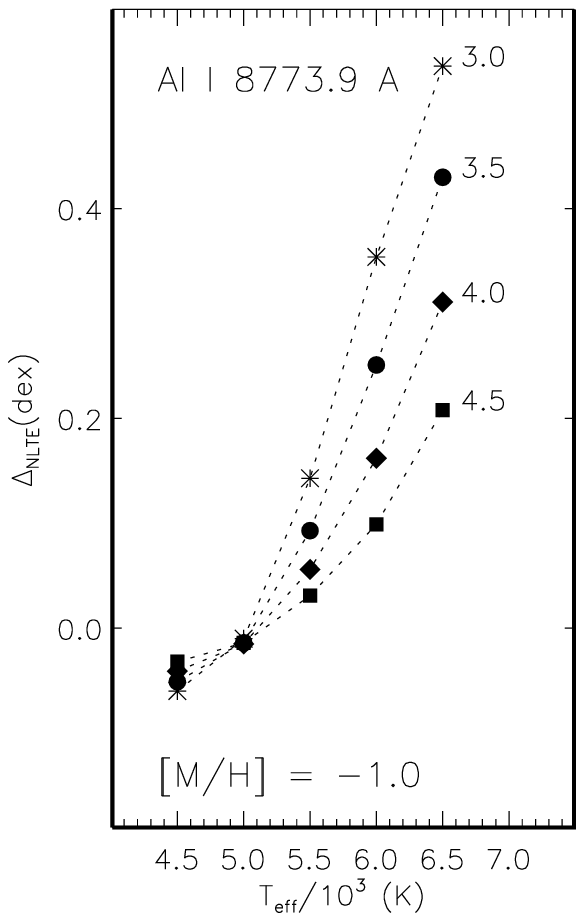}}
\caption{Non-LTE corrections for the subordinate Al~I lines as a function of the effective temperature and surface gravity in the models with [M/H] = 0 (upper row) and [M/H] =  $-1$ (lower row). Different symbols correspond to log g = 3.0 (asterisks), 3.5 (circles), 4.0 (diamonds), and 4.5 (squares). No non-LTE corrections are given for lines with  $EW$(LTE) $< 5$~m\AA. }
\label{fig:dnlte2}
\end{figure}

The non-LTE method based on accurate rates of inelastic collisions with H~I was applied to calculate
the non-LTE abundance corrections for six subordinate Al~I lines in MARCS model atmospheres
with $\Teff$ = 4500-6500~K, log g = 3.0-4.5, and [M/H] = 0 and $-1$. Even the strongest of
them, Al~I 8772 and 8773~\AA\ become unmeasurable at [M/H] $< -1$. The corrections were computed with
the LINEC code (Sakhibullin 1983), which uses the non-LTE and LTE level populations computed in the
DETAIL code. By the non-LTE correction we mean the change in abundance required to reproduce
the equivalent width (EW) calculated in LTE. Everywhere in our calculations, the aluminum abundance
follows the metal abundance in the model, i.e., [Al/M] = 0, and $\xi_t$ = 2~\kms\ in accordance with the value adopted in the computations of MARCS model atmospheres. The results are presented in Table~\ref{tab_dnlte} and, for the three lines of different multiplets, in Fig.~\ref{fig:dnlte2}.

In the coolest models with $\Teff$ = 4500 and 5000~K, the overionization of Al~I is indistinct in the subordinate line formation region, while the strong pumping resonance lines lead to an overpopulation
of the \ba{4s}{S}{}{} level. The overpopulation is then redistributed via collisions to the \ba{4p}{P}{\circ}{} and \ba{3d}{D}{}{} levels, which are the lower levels of the transitions being studied. As a
result, the subordinate Al~I lines are strengthened compared to LTE, and the non-LTE abundance corrections are negative, though small in absolute value ($< 0.1$~dex). The overionization of Al~I develops
and becomes stronger with rising $\Teff$, which leads to a weakening of the Al~I lines compared to LTE and to a positive $\Delta_{\rm NLTE}$. In metal-deficient models, the overionization of Al~I is stronger than that in models with [M/H] = 0 due to the higher intensity of ionizing
ultraviolet radiation and, therefore, $\Delta_{\rm NLTE}$ is greater.

For the subordinate lines, the non-LTE corrections were calculated as the change in abundance
required to reproduce $EW$(LTE). However, such an approach is inapplicable for the Al~I 3961~\AA\ resonance line, which is located in the wing of the very strong Ca~II 3968~\AA\ line (see, for example, Fig.~\ref{fig:profiles} for the star HD~134169 with [Fe/H] = $-0.78$). How should $EW$ of the Al~I 3961~\AA\ line be specified in this case? How should the calcium abundance be specified? We performed test calculations for the 5890/4.02/$-0.78$ model atmosphere. Our best fits of the observed Al~I 3961~\AA\ line profile in the spectrum of HD~134169 give an abundance difference of 0.77~dex between non-LTE and LTE (Table~\ref{tab_stars}). In this case, we found that [Ca/Fe] = 0.3 led to a satisfactory description of
the Ca~II 3968~\AA\ wing, and such a calcium abundance is quite expectable for a star with [Fe/H] = $-0.78$. We then performed calculations with the LINEC code, in which the background opacity includes
the Ca~II 3933, 3968~\AA\ lines, along with continuous opacity sources. The non-LTE correction for Al~I 3961~\AA\ turned out to depend on the Ca abundance in the model:

\begin{enumerate}
\item $\Delta_{\rm NLTE}$ = 0.40~dex if the Ca abundance is zero, i.e., the absorption in Ca~II 3968~\AA\ is disregarded;
\item $\Delta_{\rm NLTE}$ = 0.46~dex if [Ca/Fe] = 0,
\item $\Delta_{\rm NLTE}$ = 0.60~dex if [Ca/Fe] = 0.4.
\end{enumerate}

\noindent Thus, the synthetic-spectrum method should be used to determine the abundance from the Al~I 3961~\AA\
line irrespective of whether this is done in LTE or non-LTE. Using the non-LTE corrections
calculated from the equivalent widths and with a Ca abundance that can be far from that observed in a specific star can yield an incorrect result. In order not to mislead the potential users, we do not provide any
table of non-LTE corrections for Al~I 3961~\AA\ and do not recommend to use the non-LTE corrections for
this line irrespective of the source of their origin.

\begin{figure}  
\resizebox{50mm}{!}{\includegraphics{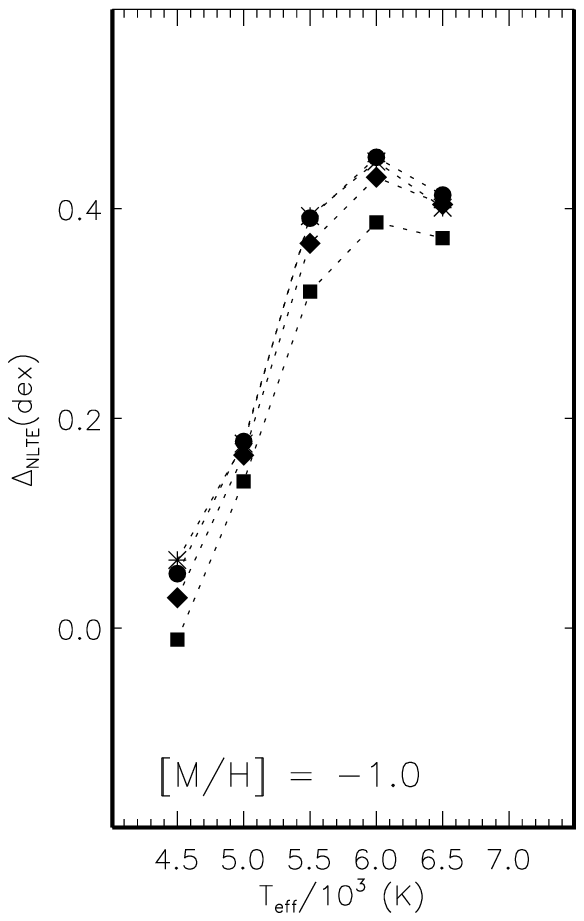}}
\resizebox{50mm}{!}{\includegraphics{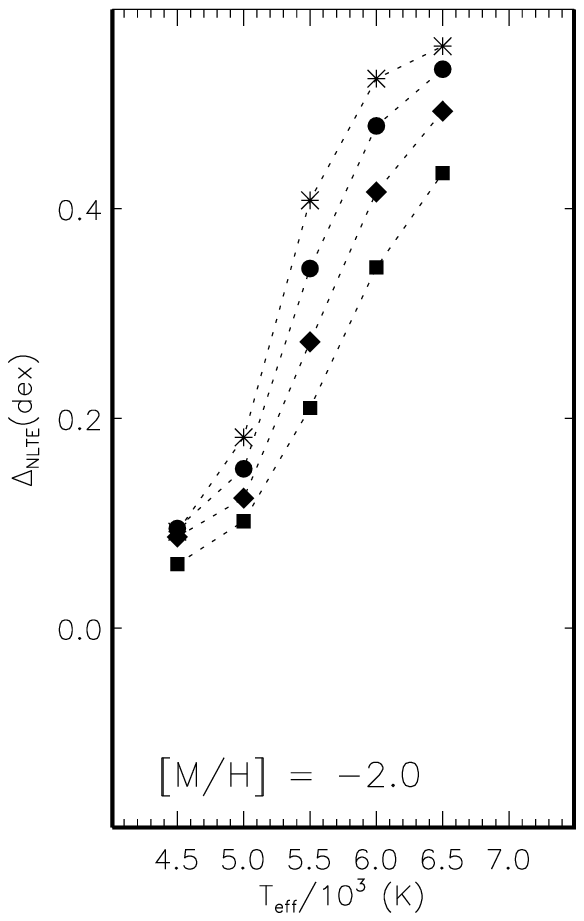}}
\resizebox{50mm}{!}{\includegraphics{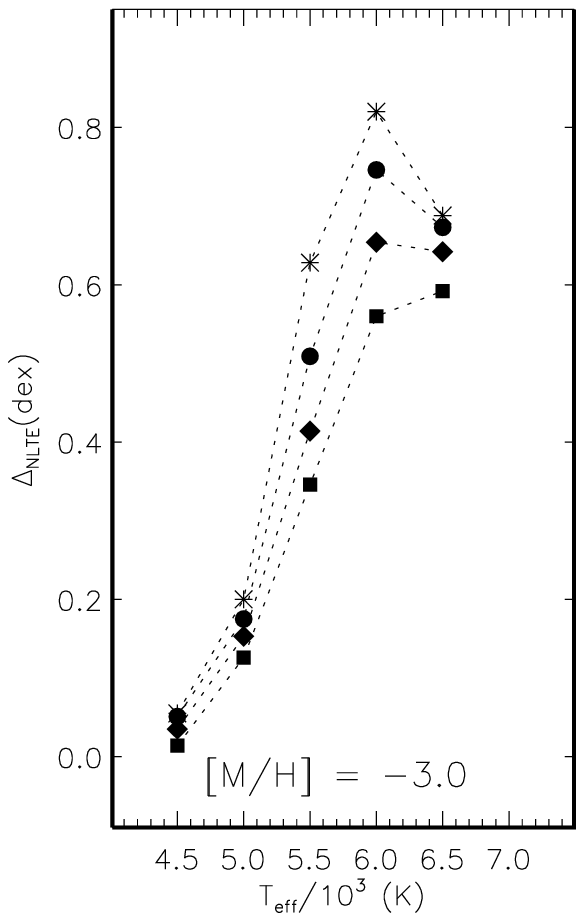}}
\caption{Non-LTE corrections for Al~I 3961~\AA\ as a function of stellar parameters. The notation is the same as that in Fig.~\ref{fig:dnlte2}. }
\label{fig:dnlte3961}
\end{figure}

To illustrate the dependence of the departures from LTE in the Al~I 3961~\AA\ line on stellar atmosphere
parameters, in Fig.~\ref{fig:dnlte3961} we present the corrections calculated by the method of equivalent widths and with a Ca abundance corresponding to the model atmosphere, more specifically, [Ca/Fe] = 0.4 for all models with [M/H] $\le -1$. Obviously, for very metal-poor stars, where the aluminum abundance can
be determined only from the resonance line, the theoretical spectra should be computed based on non-
LTE. Using LTE can lead to an underestimation of the abundance by 0.4-0.8 dex, depending on the
stellar parameters.

\begin{table}[htbp]
\caption{Non-LTE corrections (dex) for the subordinate Al~I lines as a function of the effective temperature and surface gravity in the models with [M/H] = 0 and -1. No non-LTE corrections are given for lines with $EW$(LTE) $< 5$~m\AA. }
\label{tab_dnlte}
\begin{center}
\begin{tabular}{cccccccc}
\hline\hline\noalign{\smallskip}
 $\Teff$ & log~g & \multicolumn{6}{c}{Al~I lines} \\
\cline{3-8}   
 & & 6696 & 6698 & 7835 & 7836 & 8772 & 8773 \\ 
\noalign{\smallskip}\hline \noalign{\smallskip}
\multicolumn{8}{c}{ [M/H] =  0} \\
 4500 & 3.0 &   $-$0.06 &   $-$0.04 &   $-$0.02 &   $-$0.03 &   $-$0.06 &   $-$0.08 \\
 4500 & 3.5 &   $-$0.05 &   $-$0.04 &   $-$0.02 &   $-$0.02 &   $-$0.05 &   $-$0.06 \\
 4500 & 4.0 &   $-$0.04 &   $-$0.03 &   $-$0.01 &   $-$0.01 &   $-$0.04 &   $-$0.05 \\
 4500 & 4.5 &   $-$0.03 &   $-$0.02 &   $-$0.01 &   $-$0.01 &   $-$0.02 &   $-$0.03 \\
 5000 & 3.0 &   $-$0.05 &   $-$0.04 &   $-$0.02 &   $-$0.02 &   $-$0.05 &   $-$0.06 \\
 5000 & 3.5 &   $-$0.04 &   $-$0.03 &   $-$0.01 &   $-$0.02 &   $-$0.04 &   $-$0.05 \\
 5000 & 4.0 &   $-$0.03 &   $-$0.02 &   $-$0.01 &   $-$0.01 &   $-$0.03 &   $-$0.03 \\
 5000 & 4.5 &   $-$0.02 &   $-$0.01 &   $-$0.01 &   $-$0.01 &   $-$0.02 &   $-$0.02 \\
 5500 & 3.0 &   $-$0.03 &   $-$0.02 &   $-$0.00 &   $-$0.01 &   $-$0.03 &   $-$0.04 \\
 5500 & 3.5 &   $-$0.02 &   $-$0.01 &   $-$0.00 &   $-$0.00 &   $-$0.02 &   $-$0.03 \\
 5500 & 4.0 &   $-$0.01 &   $-$0.01 &    0.00 &    0.00 &   $-$0.01 &   $-$0.02 \\
 5500 & 4.5 &   $-$0.01 &   0.00 &    0.00 &    0.00 &   $-$0.01 &   $-$0.01 \\
 6000 & 3.0 &    0.06 &    0.06 &    0.08 &    0.09 &    0.07 &    0.07 \\
 6000 & 3.5 &    0.06 &    0.06 &    0.07 &    0.07 &    0.07 &    0.07 \\
 6000 & 4.0 &    0.05 &    0.05 &    0.06 &    0.06 &    0.06 &    0.06 \\
 6000 & 4.5 &    0.04 &    0.04 &    0.04 &    0.04 &    0.04 &    0.05 \\
 6500 & 3.0 &    0.16 &    0.15 &    0.20 &    0.20 &    0.19 &    0.21 \\
 6500 & 3.5 &    0.15 &    0.14 &    0.17 &    0.17 &    0.18 &    0.19 \\
 6500 & 4.0 &    0.13 &    0.12 &    0.14 &    0.14 &    0.14 &    0.16 \\
 6500 & 4.5 &    0.10 &    0.10 &    0.10 &    0.10 &    0.11 &    0.12 \\
\multicolumn{8}{c}{ [M/H] =  $-1$} \\
 4500 & 3.0 &   $-$0.08 &   $-$0.07 &   $-$0.03 &   $-$0.03 &   $-$0.05 &   $-$0.06 \\
 4500 & 3.5 &   $-$0.06 &   $-$0.06 &   $-$0.02 &   $-$0.02 &   $-$0.05 &   $-$0.05 \\
 4500 & 4.0 &   $-$0.05 &   $-$0.05 &   $-$0.02 &   $-$0.02 &   $-$0.04 &   $-$0.04 \\
 4500 & 4.5 &   $-$0.04 &   $-$0.03 &   $-$0.01 &   $-$0.01 &   $-$0.03 &   $-$0.03 \\
 5000 & 3.0 &   $-$0.02 &   $-$0.02 &    0.02 &    0.01 &   $-$0.01 &   $-$0.01 \\
 5000 & 3.5 &   $-$0.02 &   $-$0.02 &    0.01 &    0.01 &   $-$0.01 &   $-$0.01 \\
 5000 & 4.0 &   $-$0.02 &   $-$0.01 &    0.00 &    0.00 &   $-$0.01 &   $-$0.01 \\
 5000 & 4.5 &   $-$0.01 &   $-$0.01 &   0.00 &   0.00 &   $-$0.01 &   $-$0.01 \\
 5500 & 3.0 &    0.15 &         &         &    0.17 &    0.14 &    0.14 \\
 5500 & 3.5 &    0.11 &         &         &    0.11 &    0.09 &    0.09 \\
 5500 & 4.0 &    0.08 &         &         &    0.07 &    0.05 &    0.06 \\
 5500 & 4.5 &    0.05 &         &         &    0.04 &    0.03 &    0.03 \\
 6000 & 3.0 &         &         &         &         &    0.34 &    0.35 \\
 6000 & 3.5 &         &         &         &         &    0.24 &    0.25 \\
 6000 & 4.0 &         &         &         &         &    0.16 &    0.16 \\
 6000 & 4.5 &         &         &         &         &    0.10 &    0.10 \\
 6500 & 3.0 &         &         &         &         &    0.51 &    0.54 \\
 6500 & 3.5 &         &         &         &         &    0.41 &    0.43 \\
 6500 & 4.0 &         &         &         &         &    0.30 &    0.31 \\
 6500 & 4.5 &         &         &         &         &    0.20 &    0.21 \\
\noalign{\smallskip}\hline 
\end{tabular}
\end{center}
\end{table} 

\section{CONCLUSIONS}\label{sect:conclusions}

We performed non-LTE calculations for Al~I and Si~I lines using accurate rates of collisions with
H~I from Belyaev (2013) and Belyaev et al. (2014) for models corresponding to the atmospheres of
F-G-K stars. We showed that for b-b transitions with $E_{\rm ij} \le 2$~eV, the collisions with H~I are
equally important as the collisions with electrons.
For Si~I there also exist many transitions with $E_{\rm ij} > 3$~eV, where $C_{\rm H} \simeq C_{\rm e}$. The rates of ion-pair production Al~I($nl$)+H~I $\rightarrow$ Al~II($3s^2$~$^1$S) + H$^-$ and Si~I($nl$)+H~I $\rightarrow$ Si~II($3s^2$~$^1$S) + H$^-$ are higher than the electron-impact ionization rates by two or three
orders of magnitude for levels with an ionization energy $\chi_{\rm i} < 3.5$~eV. Owing to their large cross sections, the charge exchange processes provide a close coupling of highly excited levels of neutral atoms (Al~I and Si~I) with the ground state of ions (Al~II and Si~II) and a reduction of the departures from LTE
in their populations compared to the calculations that take into account the collisions only with electrons.
For Si~I including the collisions with H~I leads to the establishment of TE populations in the line formation region even in hot metal-poor models. As a consequence, the departures from LTE in Si~I lines are minor. However, including the collisions with H~I barely affects the population of the Al~I ground state, because the
radiative rates in all the transitions that couple the ground state with the excited levels and with Al~II exceed the collisional rates. 

Our calculations confirm the previous findings of Baumueller and Gehren (1996), Andrievsky et al. (2008), and Menzhevitski et al. (2012) that are relevant to overionization of Al~I in the spectral line formation region and a weakening of lines compared to the LTE case. However, our results predict very different magnitude of the non-LTE effects, especially for the subordinate lines, compared with that from the previous studies. 

For three mildly metal-poor dwarf stars, for which there are reliable determinations of their atmospheric
parameters and high-quality observations in a wide spectral range, the aluminum abundance
was determined from seven Al~I lines in different models of their formation. Under the LTE assumption and in non-LTE calculations including the collisions only with electrons, the Al~I 3961~\AA\ resonance line
gives a systematically lower abundance compared with that from the subordinate lines, by 0.25-0.45~dex. The difference for each star is removed by taking into account the collisions with hydrogen atoms, and the rms error of the abundance derived from all seven Al~I lines decreases by a factor of 1.5-3 compared to the LTE
analysis. Thus, the collisions with H~I have to be taken into account in non-LTE calculations for Al~I. Using the rate coefficients from Belyaev (2013) plays a key role in increasing the accuracy of the derived stellar aluminum abundances.

The developed method was applied to calculate the non-LTE abundance corrections for six subordinate
lines in a grid of model atmospheres with $\Teff$ = 4500-6500~K, log~g = 3.0-4.5, and [M/H] = 0, $-1$, $-2$, and $-3$. The departures from LTE grow with towards higher effective temperature, lower metal abundance,
and lower surface gravity. For the Al~I 3961~\AA\ resonance line, we do not recommend to use the non-LTE corrections. The abundance from this line should be determined by the synthetic-spectrum
method irrespective of which approach, LTE or non-LTE, is applied.

We hope that progress in the theory will allow accurate data to be obtained in future for inelastic
collisions with H~I atoms for O~I, Fe~I, Ti~I, and other atoms that play an important role in studying
the physics of stellar atmospheres.

{\it Acknowledgements.} We are grateful to Thomas Gehren and Huawei Zhang, who provided the observed stellar spectra. This work was supported in part by the Russian Foundation for Basic Research (project nos. 15-02-06046
and 13-03-00163). L.I. Mashonkina is grateful to the Swiss National Science Foundation (SCOPES
grant no. IZ73Z0-152485) for a partial support of this study.


\begin{thebibliography}{99}

\bibitem{2008A&A...481..481A}
S.~M. Andrievsky, M.~Spite, S.~A. Korotin, F.~Spite, P.~Bonifacio, R.~Cayrel, V.~Hill, P.~Fran{\c c}ois, Astron. Astrophys., \textbf{481}, 481 (2008);
\bibitem{1995MNRAS.276..859A}
S.~D.~Anstee, B.~J.~O'Mara, MNRAS, \textbf{276}, 859 (1995);
\bibitem{Asplund2005ARAA}
M. Asplund, Ann. Rev. Astron. Astrophys., \textbf{43}, 481 (2005);
\bibitem{1998MNRAS.296.1057B}
P.~S. Barklem, B.~J. O'Mara, J.~E.~Ross, MNRAS, \textbf{296}, 1057 (1998);
\bibitem{Barklem2011_hyd}
P.~S. Barklem, A.~K. Belyaev, M. Guitou, et al., Astron. Astrophys., \textbf{530}, A94 (2011);
\bibitem{Barklem2012_mg}
P.~S. Barklem, A.~K. Belyaev, A. Spielfiedel, et al., Astron. Astrophys., \textbf{541}, A80 (2012);
\bibitem{detail}
K. Butler, J. Giddings, Newsletter on Analysis of Astronomical Spectra 9, University of London, \textbf{723}, (1985);
\bibitem{Baumueller_al1}
D.~Baumueller, T.~Gehren, Astron. Astrophys., \textbf{307}, 961 (1996);
\bibitem{Belyaev2013_Al}
A.~K. Belyaev, Astron. Astrophys., \textbf{560}, A60 (2013);
\bibitem{Belyaev2015_He}
A.~K. Belyaev, Phys. Rev., \textbf{A91}, 062709 (2015);
\bibitem{Belyaev2003_Li}
A.~K. Belyaev,  P.~S. Barklem, Phys. Rev., \textbf{A68}, 062703 (2003);
\bibitem{belyaev2010_na}
A.~K. Belyaev, P.~S. Barklem, A.~S. Dickinson, F.~X. Gad\'ea, Phys. Rev., \textbf{A81}, 032706 (2010);
\bibitem{Belyaev2012_Mg}
A.~K. Belyaev, P.~S. Barklem, A.~Spielfiedel, et al., Phys. Rev., \textbf{A85}, 032704
(2012);
\bibitem{belyaev99_na}
A.~K. Belyaev, J.~Grosser, J.~Hahne, T.~Menzel, Phys.Rev., \textbf{A60}, 2151 (1999);
\bibitem{Belyaev2014_Si}
A.~K. Belyaev, S.~A. Yakovleva, P.~S. Barklem, Astron. Astrophys., \textbf{572}, A103 (2014);
\bibitem{drawin68}
H.~W. Drawin, Z. Physik, \textbf{211}, 404 (1968);
\bibitem{fleck91}
I.~Fleck, J. Grosser, A. Schnecke, et al., Journal of Physics B: Atomic Molecular Physics, \textbf{24}, 4017 (1991);
\bibitem{Gehren1975}
T. Gehren,  Astron. Astrophys., \textbf{38}, 289 (1975);
\bibitem{Gustafssonetal:2008}
B.~Gustafsson, B.~Edvardsson, K.~Eriksson, et al., Astron. Astrophys., \textbf{486}, 951 (2008)
\bibitem{Holweger1996}
H. Holweger, Physica Scripta, \textbf{T65}, 151 (1996);
\bibitem{vald}
F. Kupka, N.~E. Piskunov, T.~A. Ryabchikova, et~al., Astron. Astrophys. Suppl., \textbf{138}, 119 (1999) - VALD;
\bibitem{kurucz2005}
R. Kurucz, Memorie della Societa Astronomica Italiana Supplementi, \textbf{8}, 189 (2005);
\bibitem{Lambert1993}
D. Lambert, Physica Scripta, \textbf{T47}, 186 (1993);
\bibitem{Lodders2009}
K. Lodders, H. Palme, \& H.-P. Gail, Landolt-B$\rm\ddot{o}$rnstein, New Ser., Astron and Astrophys., Springer Verlang, Berlin (2009);
\bibitem{Mashonkina2009}
L. Mashonkina, Physica Scripta, \textbf{134}, id. 014004 (2009);
\bibitem{Mashonkina2013_mg}
L. Mashonkina,  Astron. Astrophys., \textbf{550}, A28 (2013);
\bibitem{Mashonkina_review2013}
L. Mashonkina, IAU Symposium, \textbf{298}, 355 (2014);
\bibitem{Mashonkina2011}
L. Mashonkina, T. Gehren, J.-R. Shi, et al., Astron. Astrophys., \textbf{528}, А87 (2011);
\bibitem{2012AstBu..67..294M}
V.~S. Menzhevitski, V.~V. Shimansky, N.~N. Shimanskaya, Astrophysical Bulletin, \textbf{67}, 294 (2012)
\bibitem{NIST08}
Yu.~A. Ralchenko, E. Kramida, J. Reader, NIST ASD Team, NIST Atomic Spectra Database (version 3.1.5), {\tt http://physics.nist.gov/asd3} (2008);
\bibitem{Reetz}
J. K. Reetz, Diploma Thesis, Universit\"at M\"unchen (1991)
\bibitem{Sakhibullin83}
N. A. Sakhibullin, Tr. Kazan. Observ. 48, 9 (1983). 
\bibitem{Sitnova2015}
T. Sitnova, G. Zhao, L. Mashonkina, et al., Astrophys. J., \textbf{808}, 148 (2015);
\bibitem{steenbock84}
W.~Steenbock, H.~Holweger, Astron. Astrophys., \textbf{130}, 319 (1984);
\bibitem{Shi_si_sun}
J.~R. Shi, T. Gehren, K. Butler, et al., Astron. Astrophys., \textbf{486}, 303 (2008);
\bibitem{Shi_si_stars}
J.~R. Shi, T. Gehren, L.~I. Mashonkina, G. Zhao, Astron. Astrophys., \textbf{503}, 533 (2009);
\bibitem{Reg1962}
H. van Regemorter, Astrophys. J., \textbf{136}, 906 (1962);
\bibitem{WSM}
W.~L. Wiese, M.~W. Smith, B.~M.~Miles, Atomic transition probabilities. \textbf{Vol. 2}: Sodium through Calcium. A critical data compilation, NSRDS-NBS, Washington, D.C.: US Department of Commerce, National Bureau of Standards (1969)

\end{thebibliography}
\end{document}